\documentclass[sn-mathphys,Numbered,iicol]{sn-jnl}

\usepackage{graphicx}%
\usepackage{multirow}%
\usepackage{amsmath,amssymb,amsfonts}%
\usepackage{amsthm}%
\usepackage{mathrsfs}%
\usepackage[title]{appendix}%
\usepackage{xcolor}%
\usepackage{textcomp}%
\usepackage{manyfoot}%
\usepackage{booktabs}%
\usepackage{algorithm}%
\usepackage{algorithmicx}%
\usepackage{algpseudocode}%
\usepackage{listings}%
\usepackage{cancel}
\usepackage{bigints}

\begin{document}

\title[Article Title]{Introducing the Novatron, a novel mirror fusion concept}


\author[1]{\fnm{Jan} \sur{Jäderberg}}\email{jan.jaderberg@novatronfusion.com}

\author*[1]{\fnm{Katarina} \sur{Bendtz}}\email{katarina.bendtz@novatronfusion.com}
\equalcont{These authors contributed equally to this work.}
\author[1]{\fnm{Kristoffer} \sur{Lindvall}}
\equalcont{These authors contributed equally to this work.}
\author[1,2]{\fnm{Jan} \sur{Scheffel}}
\equalcont{These authors contributed equally to this work.}
\author[1]{\fnm{Rickard} \sur{Holmberg}}


\author[1]{\fnm{Per} \sur{Niva}}

\author[1]{\fnm{Robin} \sur{Dahlbäck}}

\author[1]{\fnm{Johan} \sur{Lundberg}}

\affil[1]{\orgname{Novatron Fusion Group}, \orgaddress{\street{Teknikringen 31}, \city{Stockholm}, \country{Sweden}}}
\affil[2]{\orgdiv{Electromagnetic Engineering and Fusion Science}, \orgname{KTH Royal Institute of Technology}, \orgaddress{\street{Teknikringen 31}, \city{Stockholm}, \country{Sweden}}}


\abstract{A new magnetic mirror-cusp concept is described - the Novatron - with the potential to confine compact and stable fusion plasmas. Traditionally, the major challenges for open field line designs include MHD interchange modes, drift cyclotron loss-cone (DCLC) modes, neoclassical transport, and axial losses of particles and energy. The novel magnetic field configuration features favorable curvature throughout the plasma region, suppressing interchange modes. Moreover, the Novatron is designed to be self-stabilized against DCLC modes by allowing for a large plasma to Larmor radius ratio. The vacuum magnetic field geometry is axisymmetric, mitigating neoclassical transport. The Novatron features a high mirror ratio, providing strong magnetic confinement and suppressed axial losses. This paper describes the fundamental magnetic field topology and outlines the design of the magnet system. MHD interchange stability of anisotropic low-$\beta$ equilibria is demonstrated by derivation of two novel criteria, based on anisotropic ideal MHD and the Chew-Goldberger-Low model, and numerical computation in Novatron geometry. The Novatron design is also placed into a historic context by summarizing challenges faced by both previous and more current mirror/cusp concepts. 
}

\keywords{Fusion plasma, plasma devices, open field line design, axisymmetric magnetic mirror, MHD stability}



\maketitle

\section{Introduction}\label{sec:intro}

The Novatron is an innovative fusion concept with a mirror-cusp magnetic topology addressing several longstanding challenges in open field line fusion research. It provides the interchange stabilizing features of the successful Yin-Yang coils~\cite{Moir}, but in an axisymmetric system with a geometry that has a much higher mirror ratio and larger plasma volume than the classic cusp. The Novatron design has a low aspect ratio --- the ratio between length and diameter (short and wide) --- enabling many Larmor radii across the plasma radius. These design choices and the challenges they address are summarized in Table~\ref{tab:solutions}. The challenges will be described further in the remainder of this section.


\begin{table}[h]
\caption{Critical issues for mirror machines and the Novatron approach to address them. Here $r_p$ here denotes the plasma radius and $r_L$ the Larmor radius.}\label{tab1}%
\begin{tabular}{@{}ll@{}}
\toprule
\textbf{Challenge} & \textbf{Novatron approach}  \\
\midrule
Interchange instability    & Favorable curvature  \\
DCLC instability   & High $r_p$/$r_L$ ratio  \\
Neoclassical transport   & Axial symmetry   \\
Axial losses   & High mirror ratio \\
\botrule \label{tab:solutions}
\end{tabular}
\end{table}

Since the beginning of fusion research, several promising reactor designs have been proposed and studied. Two principal branches of magnetic confinement emerged during this time: designs with closed field lines, such as tokamaks and stellarators, and designs with open field lines, including mirror machines and cusps. By the mid-1980s, both branches made steady progress~\cite{El-Guebaly}, but major funding bodies decided to narrow focus to the tokamak, which at the time was outperforming mirror machines. However, the past four decades of mirror machine research have created significant breakthroughs~\cite{Simonen2008}, paving the way for mirrors and cusps to potentially become viable, cost-efficient devices.

A challenge for mirror confinement is the interchange (flute) instabilities occurring at plasma surfaces with “unfavorable” curvature~\cite{TellerB}. Unfavorable curvature, as opposed to favorable curvature, is defined as a decrease in magnetic field strength in the direction of decreasing plasma pressure. This corresponds to convex magnetic field lines as seen from outside the plasma volume. The unfavorable curvature, in a field line averaged sense, is indeed an intrinsic problem for configurations based on the classic magnetic mirror design. Favorable curvature has previously been achieved by minimum-B configurations, but at the expense of non-axisymmetric geometry, causing extensive neoclassical (radial) particle losses. The Novatron, by contrast, is axisymmetric and exhibits favorable curvature.

Another problem for mirror designs is axial particle loss --- loss through the mirror throats --- that occurs for particles with a too high axial velocity as compared to radial velocity. The region in velocity space corresponding to plasma leaking out is referred to as the loss cone~\cite{Ryutov1988}. The Novatron has a high mirror ratio, inhibiting extensive axial losses.

Drift cyclotron loss-cone (DCLC) modes are electrostatic microinstabilities with frequencies near the ion cyclotron frequency and wavelengths smaller than the ion Larmor radius. They typically arise because of gradients in the velocity distribution ~\cite{Post1966} or where the density gradient in the radial direction exceeds a critical value. If present, DCLC modes give rise to excessive particle and energy losses. A common way for DCLC modes to form is by fast electrons escaping from the end of the plasma through the loss cone, resulting in the formation of a plasma potential which is negative outside the mirrors and positive inside the central cell, commonly referred to as ambipolar potential~\cite{El-Guebaly}. The positive plasma potential expels low energy ions, causing a void in the ion velocity distribution, an “ambipolar hole". These anisotropic, non-Maxwellian ion distributions can drive DCLC micro-instabilities~\cite{Kotelnikov:2017}. The large $r_P/r_L$ ratio of the Novatron promotes stability against DCLC modes.

A first, 3.8 m tall and 1.8 m wide, prototype of the Novatron concept, the N1 experiment, is currently under completion at the Alfvén Laboratory at the KTH Royal Institute of Technology, Stockholm, Sweden. The main goal of the N1 experiment is to validate the intrinsic interchange stability of the Novatron magnetic topology. Diagnostics currently under preparation include Langmuir probes, diamagnetic loops, visual high speed cameras, photodiodes and an RF interferometer.

 The paper is organized as follows: Sec.~\ref{sec:mag_design} provides a comprehensive layout of the Novatron design. It begins by focusing on the fundamental topology and draws comparisons with classic mirrors, cusps, and mirror-cusps. Subsequently the topics of magnetic domains, aspect ratio, number of Larmor radii per plasma radius, weak field regions, mirror ratio, and expanders are discussed. Sec.~\ref{sec:history} gives a brief history of the mirror/cusp experiments at Lawrence Livermore National Laboratory (LLNL) and Nagoya University. Sec.~\ref{sec:current} presents an overview of other current mirror concepts and techniques to address the challenges for open field line configurations. Anisotropic Novatron equilibria are presented in Sec.~\ref{sec:eq_stab}, where also two novel interchange stability conditions are derived. These are subsequently evaluated numerically for the NovatronA discussion, primarily on topics to be presented in parallel papers, is given in Sec.~\ref{sec:discussion}. Conclusions are drawn in Sec.~\ref{sec:conclusion}. 

\section{The Novatron magnetic field geometry}\label{sec:mag_design}

In this section, the Novatron magnetic field geometry is presented by highlighting some of its most defining features. We begin by describing the magnetic field curvature, where comparisons to mirrors and cusps are made, followed by discussions on aspect ratio and $r_P/r_L$ ratio, weak field domains, mirror ratio and expanders.

\subsection{Axial symmetry and favorable curvature --- comparison to the classic mirror and the cusp}

\begin{figure}[h]%
\centering
\includegraphics[width=76mm]{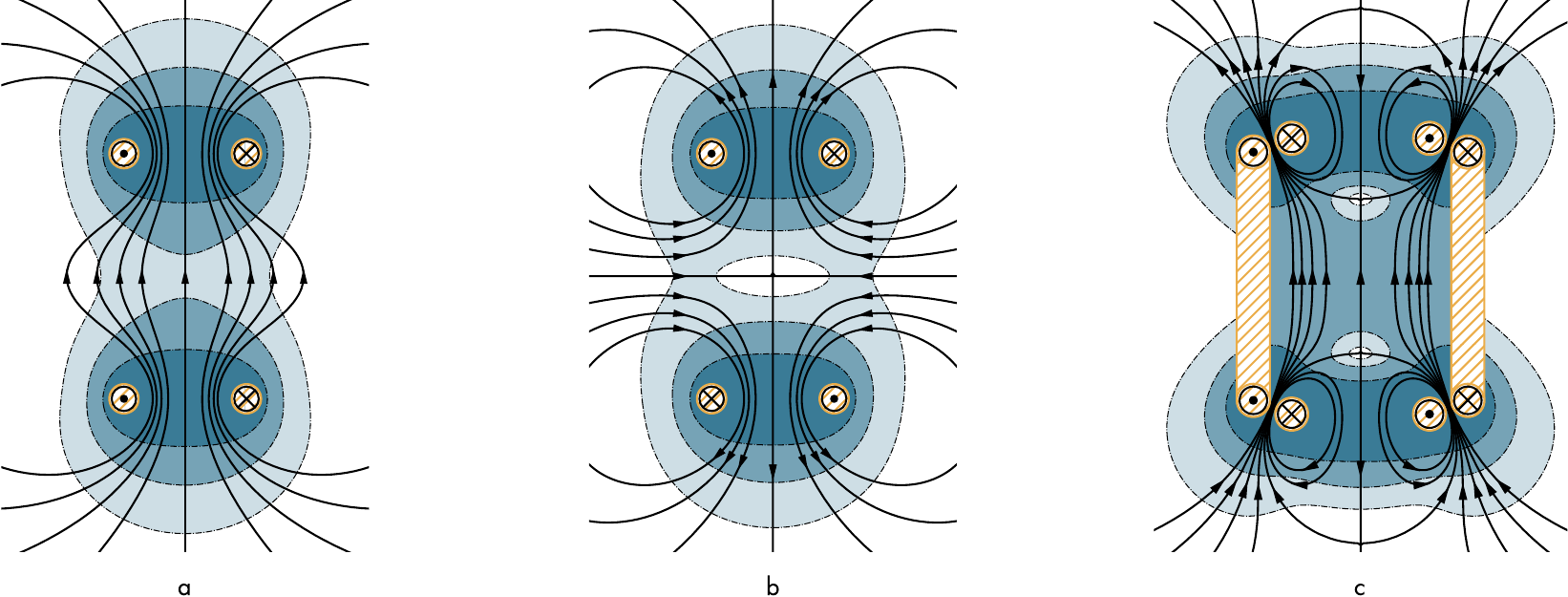}
\caption{Coils and field lines for a schematic classic mirror (a), a biconic cusp (b), and a Novatron (c). Current directions in the coils are marked with a cross (×) for current directed into the plane and a dot ($\cdot$) for current directed out of the plane of view.}\label{fig:cusp}
\end{figure}

Two basic fusion concepts within the open field line group are the classic magnetic mirror (partially unfavorable curvature) and the biconic cusp (favorable curvature). Fig.~\ref{fig:cusp} shows the fields for a classic mirror, a biconic cusp and the Novatron, side by side. The sections where the field is the strongest are called mirror throats. (For the Novatron, we refer to the regions of stronger field along the central as the chimneys). The symmetry plane (horizontal in Fig.~\ref{fig:cusp}) is the plane dividing the field in two symmetric parts. The classic magnetic mirror has a magnetic field pointing in the direction normal to the symmetry plane, while the biconic cusp has magnetic field lines tangential to the symmetry plane. Within this context, the Novatron constitutes its own category of magnetic confinement with a perpendicular magnetic field at the symmetry plane like the classic mirror, but with favorable curvature like the biconic cusp.

\begin{figure}[h]%
\centering
\includegraphics[width=76mm]{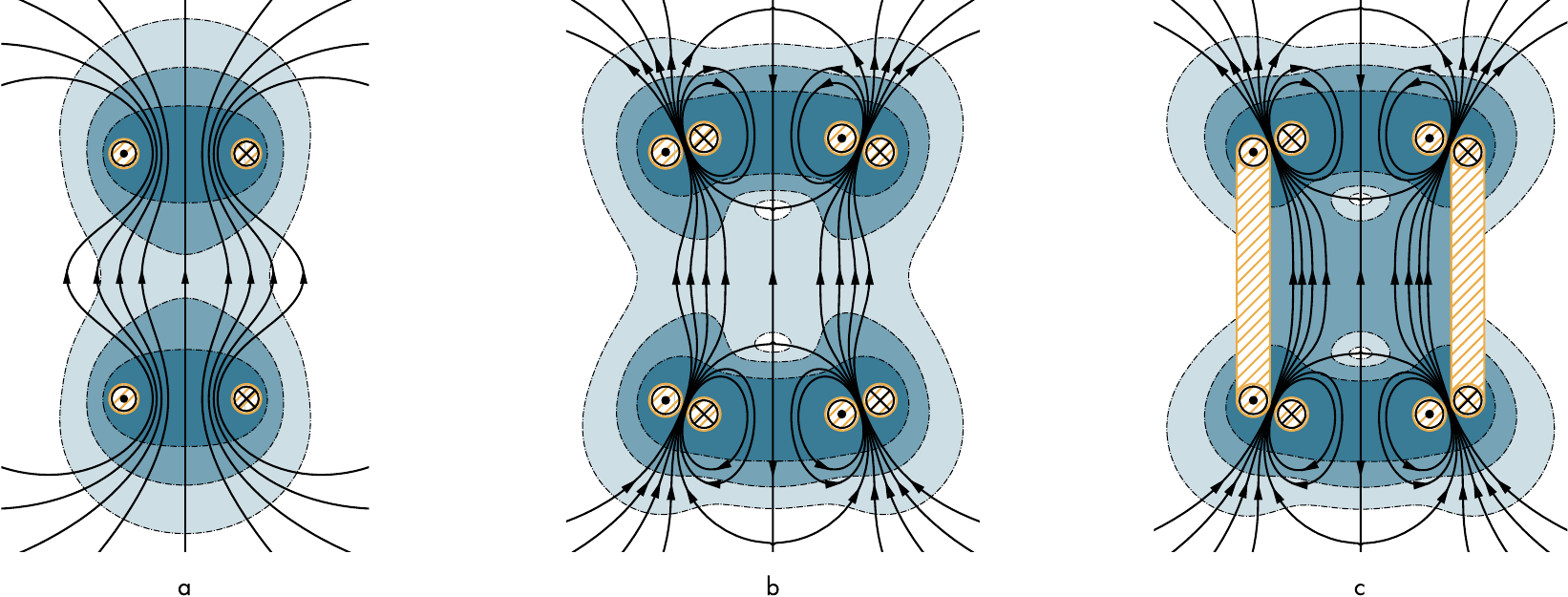}
\caption{A schematic classic mirror (a). Adding a concentric coil with opposite current (b) will give a favorable curvature of the inner field lines. A solenoid coil radially outside the plasma volume with the current in the same direction as the second coil system (c) will transform the outer field line curvature from unfavorable to favorable.}\label{fig:progress}
\end{figure}


The favorable curvature at the outer plasma surface is due to an increasing magnetic field in the outward direction. It is theoretically established~\cite{Rosenbluth1957} that such a field will provide MHD stability for an inwardly directed plasma pressure gradient. The axisymmetric design prevents neo-classical transport, that is non-collisional particle drifts in the radial direction~\cite{Fowler2017}. Such losses constituted a considerable problem for minimum-B stable configurations which had abandoned axisymmetry, like Baseball or Yin-Yang coils.

Fig.~\ref{fig:progress} helps to explain how the Novatron differs from a classic magnetic mirror. The latter (Fig.~\ref{fig:progress}a) features  a relatively large plasma volume, yet with a geometry allowing for detrimental interchange instabilities to develop owing to the unfavorable curvature in the region between the mirror throats. By adding a second magnetic coil pair located concentrically outside the first magnet pair (Fig.~\ref{fig:progress}b), and with current running in the opposite direction with respect to the first pair, the magnetic flux will be squeezed into the mirror throats, i.e. the magnetic field will be stronger in these regions. The curvature of the central field lines passing through the symmetry plane has now been transformed from unfavorable to favorable. This constitutes the most fundamental distinction between the Novatron concept and a classic mirror. 

\begin{figure}[hbt!]%
\centering
\includegraphics[width=78mm]{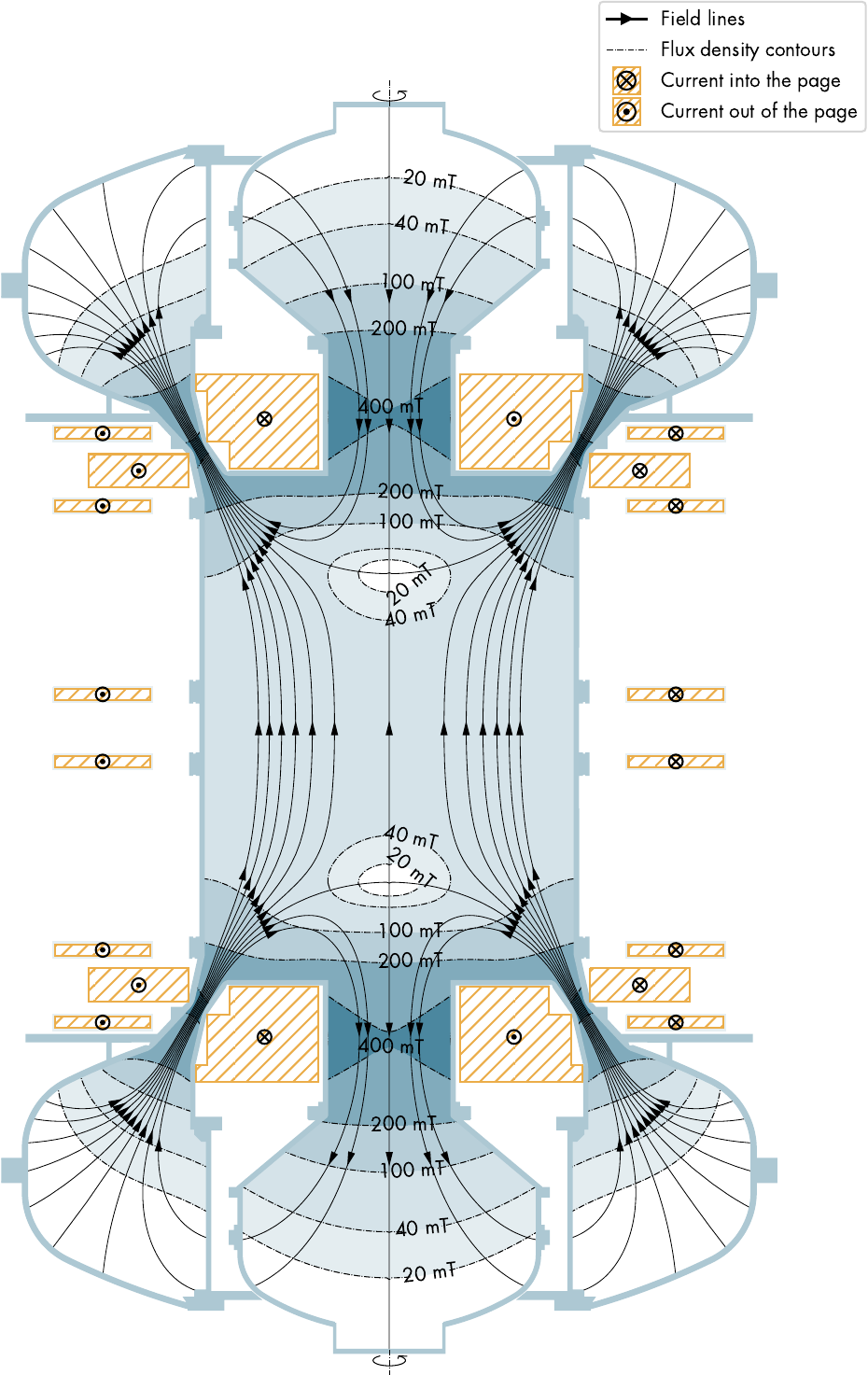}
\caption{The shape of the vacuum magnetic field lines for the Novatron topology with its characteristic features. Axisymmetry is indicated by the arrows around the central axis. A central cell and four expanders are shown. Two expanders include the central axis and are connected to the central cell by the high field strength region called the chimney. The two other expanders are donut shaped and are connected to the central cell via high field strength regions called mirror throats. All boundary flux surfaces passing through the mirror throats will have favorable curvature. The field strengths given are those of the N1 experiment, see Sec.~\ref{sec:intro}.\label{fig:N1_overview}}
\end{figure}

Adding a third magnetic coil system (Fig.~\ref{fig:progress}c) located between the two mirror throat volumes, with a current running in the same direction as the second magnet coil pair, transforms also the outer flux line curvature from unfavorable to favorable. Hence, all field lines crossing the symmetry plane and entering the mirror throats will exhibit favorable curvature and the total flux in the plasma confinement domain is increased. A Novatron magnetic field configuration, with magnetic field strength values planned for the N1 experiment, is shown in Fig.~\ref{fig:N1_overview}. 


\subsection{Magnetic domains}


\begin{figure}[h]%
\centering
\includegraphics[width=76mm]{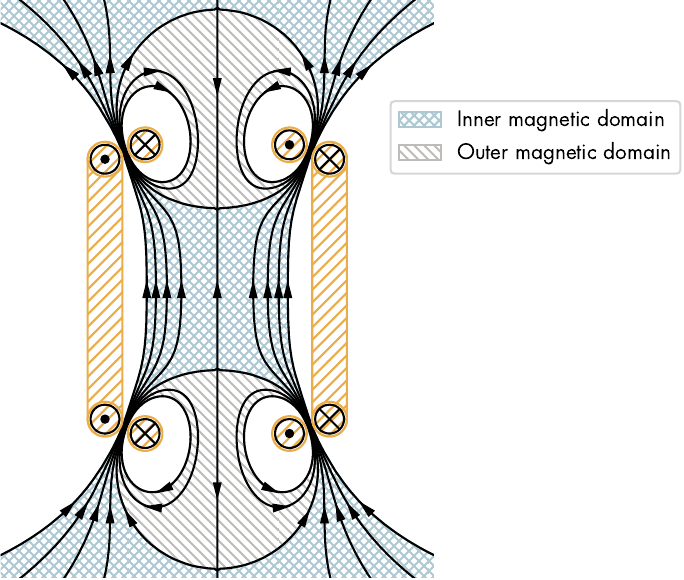}
\caption{The Novatron concept offers two distinct magnetic domains, the inner magnetic domain (bluegreen grid) and additional outer magnetic domain (gray stripes).}\label{fig:mag_domains}
\end{figure}


Adding the second magnet system (Fig.~\ref{fig:progress}b), creates magnetic separatrices, separating flux going in the clockwise direction from flux going in the counterclockwise direction. This can be seen in Fig.~\ref{fig:mag_domains}, showing the different magnetic domains of the Novatron. 

 Flux lines from the mirror throat passing through the symmetry plane extend through the corresponding mirror throat on the opposite side of the symmetry plane. This flux makes up the ``inner magnetic domain'' (bluegreen grid in Fig.~\ref{fig:mag_domains}) which is the main plasma domain. The associated favorable curvature and high mirror ratio are expected to support inner domain plasma stability. 

The Novatron concept however provides additional regions that support MHD stability
--- the ``outer magnetic domains'' (gray striped in Fig.~\ref{fig:mag_domains}). These are formed by flux passing through the mirror throat, bending back into the chimney. The chimney constitutes the region with the highest B-field in the system.

\begin{figure}[h]%
\centering
\includegraphics[width=76mm]{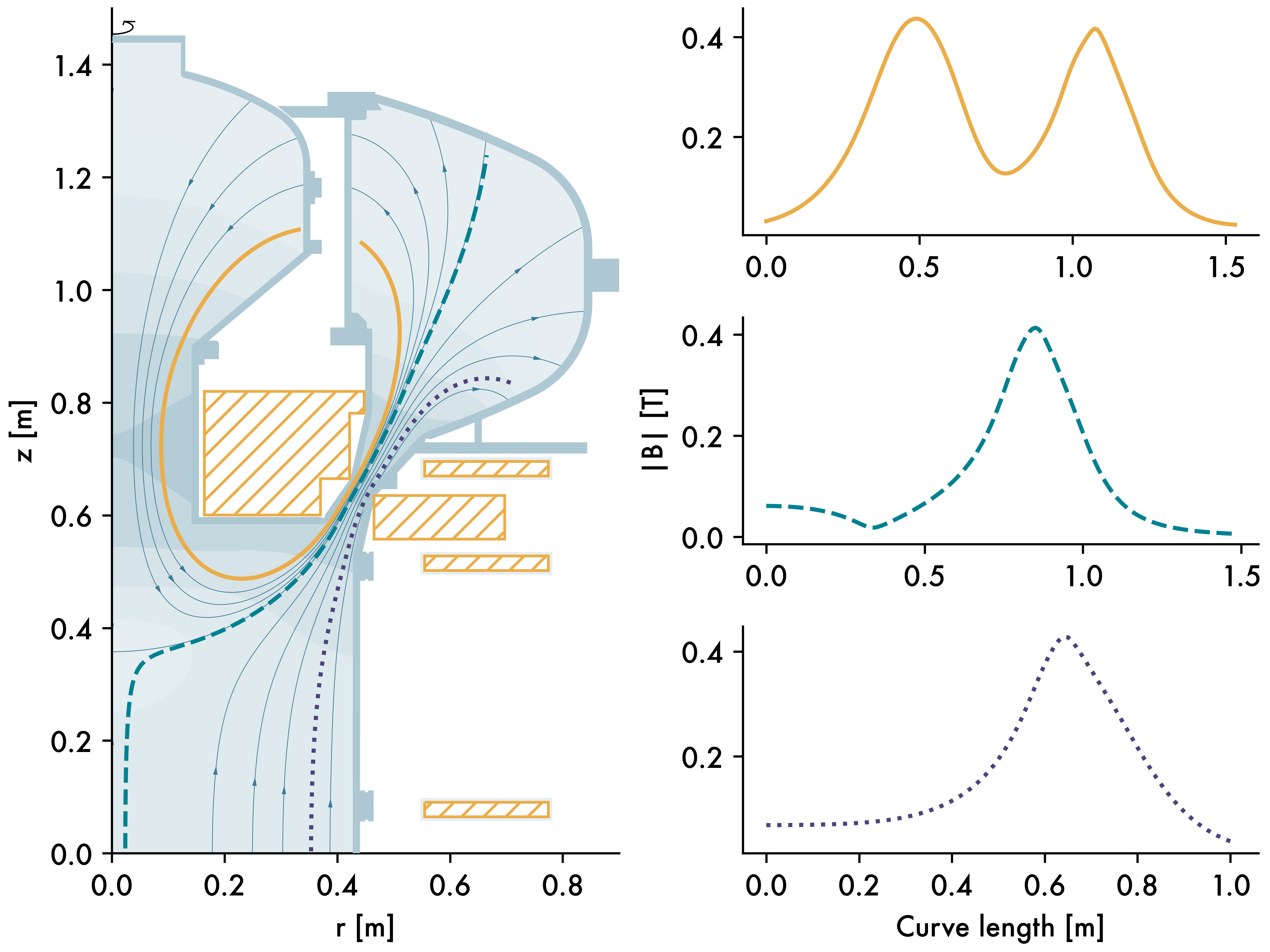}
\caption{Magnetic field strength for the N1 experiment vs.\ curve length for three field lines characteristic of the Novatron magnetic field. The top curve is the field line encircling the innermost magnet, the middle curve is a field line running within the inner magnetic domain close to the separatrix, and the bottom curve is the field line at the edge of the plasma, as marked in the figure.}\label{fig:lines}
\end{figure}

\begin{figure}[h]%
\centering
\includegraphics[width=76mm]{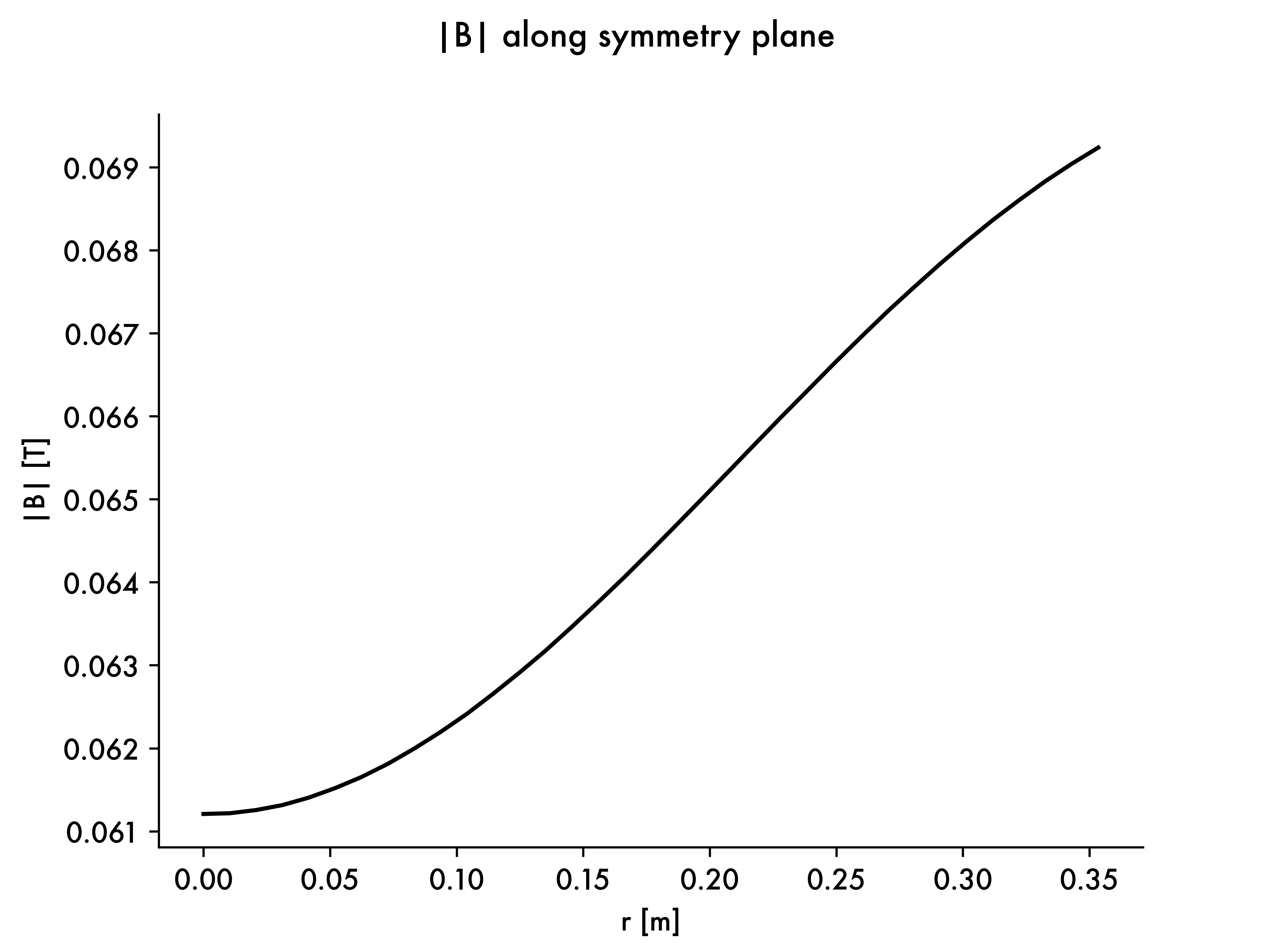}
\caption{Magnetic field strength at the symmetry plane, for the N1 experiment.} \label{fig:lines_sym_plane}
\end{figure}

Fig.~\ref{fig:lines} shows the magnetic field strength along field lines in the inner and outer magnetic domains, as well as along the separatrix, whereas Fig.~\ref{fig:lines_sym_plane} shows the magnetic field strength along the symmetry plane.


\subsection{Aspect ratio and number of Larmor radii}

The low aspect ratio (“short and wide” design) of the Novatron (see Fig.~\ref{fig:N1_overview}) supports a high ratio $r_p/r_L$ (where $r_p$ is the plasma radius in the symmetry plane and $r_L$ is the ion Larmor radius), important for DCLC stabilization~\cite{Coensgen1976}.

The DCLC modes are predicted to disappear at values of $r_p/r_L= 20-100$~\cite{El-Guebaly, Ferron1984}. As a historic comparison, the $r_p/r_L$ ratio in the 2XIIB was typically 2 (except for a large plasma radius test) resulting in DCLC mode activity limiting its performance~\cite{Hua2004}. Injected ``warm'' plasma suppressed the DCLC modes but at the cost of cooling the hot central fusion plasma~\cite{TellerA, Coensgen1972}, see Sec.~\ref{sec:LLNL}. A more current example is the WHAM experiment, planning for an $r_p/r_L$ ratio of 3--4~\cite{Endrizzi2023}. To stabilize against DCLC modes, sloshing ions and plasma rotation will instead be employed, see Sec.~\ref{sec:wham}.

The Novatron design, where DCLC modes are suppressed mainly by a large $r_p/r_L$ value, reduces the reliance on other methods. This has the potential to constitute a substantial advantage for enhanced $n \tau_E$-values in the Novatron as compared to, for example, 2XIIB.

\subsection{Domains where \texorpdfstring{$B=0$}{B=0}}\label{B=0}

At two points, where the separatrices cross the axial symmetry axis, the magnetic field is zero, as seen in Fig.~\ref{fig:N1_overview}. Charged particles passing near these points will no longer be adiabatically confined and are likely to be scattered into the loss cone and disappear from the plasma region via the separatrix. In Sec. \ref{sec:J_mu_Novatron} this matter is investigated further.

The increased leakage rates near the weak-field regions are likely to cause a hollow density profile, with its minimum on the z-axis. The density gradient in this region will then be directed radially outwards, rendering the external magnetic field curvature unfavorable. This problem is, however, resolvable as will be discussed in greater detail in Sec.~\ref{sec:interchange}.

\subsection{Mirror ratio \texorpdfstring{$R_M$}{RM}}

A core feature of the Novatron design is the high mirror ratio, crucial for suppressing axial losses. For a reactor-sized Novatron, the mirror ratio is planned to exceed 16. Even assuming the conservative $\log_{10}(R_M)$ scaling of $n \tau_E$ ~\cite{Post1987}, a high mirror ratio will contribute significantly to the confinement time. Fig.~\ref{fig:lines} shows the magnetic field strength along three different field lines.

\subsection{Expanders} \label{sec:expanders}

The expansion ratio is defined as the ratio of the B-field at the mirror throats ($B_M$) compared to the B-field at the end walls in the outer expanders (see Fig.~\ref{fig:N1_overview}). The expansion of the magnetic flux density is important to prevent cooling of the plasma stemming from secondary electrons emitted off the end plates~\cite{Soldatkina2017}. An expansion ratio exceeding the square root of the ion to electron mass ratio ($\approx 60$ for deuterium plasmas~\cite{mirnov1980}) has been shown to be necessary, notably constituting another limiting factor of previous mirrors such as the 2X experiments. The Novatron design enables a large expansion ratio, exceeding the desired expansion ratio of 60.

\section{Historic background} \label{sec:history}
As described in the previous section, the Novatron mirror-cusp concept is intended to resolve the major  difficulties that earlier mirror and cusp research faced. A brief contextualizing historic background is given in this section.

\subsection{Mirrors at the Lawrence Livermore National Laboratory (LLNL)} \label{sec:LLNL}

In the early 1950s, an ambitious magnetic mirror fusion research program~\cite{Bishop1958}, led by Richard F. Post, was launched at the LLNL.\@ The research program included the three consecutive 2X machines~\cite{Coensgen1972}, where Baseball coils were first used to stabilize against interchange modes. The second experiment, called the 2XII machine~\cite{Coensgen1972}, equipped with the more compact Yin-Yang coils~\cite{Moir}, yielding higher mirror ratio, demonstrated an improved suppression of interchange instabilities and increased confinement time. The upgraded version, 2XIIB~\cite{Coensgen1976} (B for beams), was amended to include 12 neutral beam modules of 20\,kV.\@ The neutral beams raised the temperature, and the plasma density reached above $10^{20}\,$m$^{-3}$, but the confinement time was limited by DCLC instabilities and the inability to reach fusion relevant temperatures. A solution involving injecting a stream of plasma near the mirrors to control the DCLC instabilities was implemented~\cite{Post1987}.

In 1975, 2XIIB became the first fusion experiment to reach 100 million degrees~\cite{Fowler1997}, three years before the Princeton Large Torus (PLT) Tokamak. To this day, 2XIIB holds the record for mirror machines in terms of ion temperature, triple product, and high $\beta$. The 2XIIB experiment proved the qualitative feasibility of mirror fusion: instabilities could be adequately controlled and a dense plasma at thermonuclear temperatures could be created and sustained~\cite{Kesner1983}.

However, further enhancement of the energy confinement time $\tau_E$ and of the triple product encountered several limiting factors: DCLC modes, neoclassical transport and structural problems. The flow of low-energy plasma which was necessary to suppress DCLC modes, as mentioned above, had a cooling effect, as noted by Refs.~\cite{Coensgen1972, Fowler2017, Baker1984}. All the minimum-B configurations, i.e.\ with field line averaged favorable curvature, such as Ioffe bars, Yin-Yang and Baseball coils, achieved the stabilizing minimum-B property by abandoning axisymmetry, thus introducing neoclassical, radial transport~\cite{TellerB}. There were structural and mechanical complications in designing the Yin-Yang coils. For example, Yin-Yang coils are more difficult to wind, and magnetic forces act with straightening forces on any non circular coil, placing them in tension, as discussed by Teller~\cite{TellerB}.

\subsection{Mirror-cusps at Nagoya University}\label{sec:RFC}

 In the 1960s a research program was initiated at the Institute of Plasma Physics at Nagoya University, Japan, eventually demonstrating the success of Radio Frequency (RF) ponderomotive plugging, a technique to confine the ions, see Sec.~\ref{sec:techniques}, in cusps with densities of $10^{18} $m$^{-3}$ and a temperature of 0.4\,keV~\cite{Sato1985}.  A series of experiments were carried out. The first experiment was the  TPD-II~\cite{Hiroe1975} and the series was finalized by the RFC-XX-M~\cite{Okamura1984}. While the TPD experiments consisted of a single cusp, the RFC-XX experiments had a central cell (which could be run both in a configuration with a uniform field as well as with a mirror field) with the cusps placed at the ends, introducing anchor functionality (see Sec.~\ref{sec:techniques}). RFC-XX-M proved to be stable under a range of plasma conditions despite the fact that the inside of the plasma had unfavorable curvature~\cite{Baker1984}.

The RFC-XX magnetic field, though bearing some resemblance to the Novatron in that it has favorable (average) curvature, is different in several respects. The most distinguishing difference is that the RFC-XX has mirror throats perpendicular to the central axis. By increasing the inclination angle with respect to the central axis, the Novatron can reach a substantially higher mirror ratio (RFC-XX\cite{Teruyuki:1983} had only 2.1 in the uniform field configuration) for a larger plasma volume, thus allowing for a larger region of magnetically confined plasma. Another important difference, the Novatron is equipped with expanders so as to avoid secondary emission electrons from the end walls to cool the plasma, see Sec.~\ref{sec:current}.

\section{Other mirror physics development} \label{sec:current}

To put the Novatron approach to the challenges of mirror physics in context, this section delineates how other mirror concepts, both past and present, have addressed these problems. 

\subsection{Techniques to suppress interchange modes, DCLC modes, and axial losses}\label{sec:techniques}

It is generally accepted that the confinement characteristics of a classic mirror, simple mirror device are insufficient for achieving Q factors (ratio of produced fusion power to heating power) of relevance for a full-scale reactor. Current mirror concepts address the confinement and stability problems related to mirrors in various ways, with a general focus on axisymmetric designs with unfavorable curvature in the central cell and where the interchange instability is (planned to be) stabilized by other means. One approach is so called vortex stabilization~\cite{Beklemishev2010} by which the plasma is stabilized by rotation around the central axis. Another approach is anchor cells~\cite{Ryutov2011}, regions outside the central cell with favorable curvature making up for the unfavorable curvature in the central cell.  According to the criteria derived by Rosenbluth \& Longmire~\cite{Rosenbluth1957} favorable curvature at one section of the field line can under certain conditions make up for unfavorable curvature at another section. A third technique, line-tying, makes use of conducting plates at the end of the plasma region allowing for electrons to flow between field lines and thus short-circuit unstable flux tubes~\cite{Katanuma2013}.

Potential DCLC instabilities, in turn, can be addressed by filling the ambipolar hole~\cite{Ryutov1988}. The ambipolar hole can be filled by injecting ``warm'' (and isotropic) plasma ions, or by a technique commonly referred to as sloshing ions~\cite{Fowler2022}. Ions injected\footnote{Sloshing ions can also be created using neutral beams directed at an angle to the field lines.} at an angle with respect to the field lines cause ions to bounce between mirrors, making it possible to create a potential at the (tunable) turning point, thereby trapping warm thermal ions~\cite{Simonen1983}. Another, more recent approach is to use the centrifugal potential associated with plasma rotation to trap low-energy ions~\cite{Endrizzi2023}. 

Expanders are cells placed outside the mirror throats, where the magnetic field is allowed to “fan out” (see Fig.~\ref{fig:N1_overview}). As mentioned in Sec.~\ref{sec:expanders}, by diluting the influx of plasma on the end plates, secondary emission electrons from the end plates which potentially cool the plasma, are suppressed~\cite{Ryutov2004}. The resulting favorable curvature of the field in the expanders can also be used to stabilize against interchange by the same principle as for anchor cells~\cite{Ryutov2011}. Expanders additionally offer the opportunity to provide a radial electric field using biased collectors used for vortex stabilization in the central cell. Both interchange stabilization and vortex stabilization using expanders rely on a certain degree of plasma flow between the central cell and the expanders. For the vortex confinement, the plasma needs to be in electrical contact with the absorber electrodes in the expanders.

To confine the plasma axially, a high mirror ratio can be supplemented with the introduction of ambipolar electrostatic fields in so-called tandem cells, magnetically plugged regions extending the central cell in the axial direction. This is commonly referred to as electrostatic or ambipolar end-plugging and is used for confining the central cell ions. Electrostatic end-plugging can be augmented by a modification of the electrostatic potential in the region between the central and tandem cells, commonly referred to as a thermal barrier, extending the confinement to also include electrons. Radio Frequency (RF) plugging is another approach to prevent axial losses by making use of the ponderomotive force acting on the plasma particles when an oscillating electrical field is applied. 

\subsubsection{Gas Dynamic Trap (GDT) and related experiments}

The Gas Dynamic Trap (GDT)~\cite{Ivanov2013} developed at the Budker Institute, Novosibirsk, Russia, is a classic axisymmetric mirror. The GDT has a long mirror-to-mirror distance, exceeding the effective mean free path of ion-ion scattering into the loss cone, and a high mirror ratio, resulting in an isotropic and Maxwellian plasma. The central cell is supplemented by expanders. To further stabilize against interchange modes, several methods have been researched including line tying, anchor cells as well as vortex confinement using the expanders~\cite{Ivanov2016}. The vortex confinement method currently used for stabilizing against interchange modes requires a certain axial loss for the plasma to be in direct electrical contact with the absorber electrodes~\cite{Simonen2010}. The axial loss constitutes a challenge for reactor size scaling~\cite{BagryanskyIPS} and ambipolar electrostatic barriers in the expanders are currently being investigated to address the axial loss problem~\cite{SoldatkinaIPS}. Alternative remedies including several multi-mirrors (for example the GDMT experiment~\cite{Skovorodin2023}) and/or helical confinement extensions, the SMOLA  experiment~\cite{Sudnikov2017}) of the GDT concept have been, or are, under investigation. Regarding DCLC, injection of warm plasma with an isotropic distribution has been employed where a few percent of isotropic ions are sufficient for suppression. However, this method reduced the theoretical Q factor from 1 to 0.04 owing to axial losses of the warm ions~\cite{BagryanskyIPS}. There is to date a lack of agreement between theoretical and experimental accounts of DCLC in the GDT experiment~\cite{Kotelnikov2017}.

\subsubsection{GAMMA 10}

The GAMMA 10 experiment~\cite{Inutake} at the University of Tsukuba is the world's largest existing mirror machine, measuring 27\,m in the axial direction. It has an axisymmetric central cell mirror and is equipped with tandem cells used for electrostatic and thermal end plugging, which has been shown to suppress axial losses successfully~\cite{Inutake}. To stabilize against interchange modes, the design includes Baseball coil anchor cells, but interchange instabilities have been experimentally observed~\cite{Minami2002}. The sloshing ions approach has been used to stabilize against DCLC~\cite{Yatsu1984}. A density limit of $2 \times 10^{18}$ m$^{-3}$ has been observed~\cite{Kawamori2002} owing to challenges in Ion Cyclotron Radio Frequency (ICRF) heating for high density plasmas. This difficulty constitutes an active area of research for the experiment. Another concern for the GAMMA 10 is the low electron temperature which leads to cooling of the ions. In recent years, the GAMMA 10 research program has primarily revolved around investigations relevant to the large international tokamak experiments ITER and DEMO.\@

\subsubsection{The Wisconsin HTS Axisymmetric Mirror (WHAM)}\label{sec:wham}

Inspired by GDT, the WHAM experiment~\cite{Endrizzi2023} at the University of Wisconsin-Madison, USA, is conceptualized as a classic axisymmetric mirror with high mirror ratio (up to 16). Several experimental approaches are planned for stabilization of interchange modes including vortex stabilization. To achieve the high mirror ratio and high $\beta$, a strong (17 T) magnetic field will be induced from high temperature superconducting (HTS) magnets. Sloshing ions and centrifugal potential from plasma rotation are planned approaches for suppressing DCLC modes. An active area of research complementing the WHAM experiment is in prediction modeling. The modeling encompasses, among other things, stability analysis of anisotropic MHD equilibrium profiles, Fokker-Planck simulations, kinetic instability analysis, and particle confinement time estimates.

\section{Novatron equilibrium and stability}\label{sec:eq_stab}

A Novatron fusion plasma, being magnetically confined in an open-ended geometry, cannot be assumed to be isotropic. For confinement, the parallel pressure must tend to zero towards the mirror throats, thus $p_{\perp} > p_{\parallel}$~\cite{Taylor:1963}. This is inconsistent with the MHD equilibrium equations for isotropic plasmas where the plasma pressure is constant along the field lines. We therefore investigate the interchange stability assuming an anisotropic plasma.  

As was discussed Sec.~\ref{B=0}, the Novatron has two $B=0$ regions. Since adiabatic particle motion will be violated in these regions, it is expected that the axial plasma leakage will be more extensive here as compared to the rest of the volume, leading to lower plasma density and pressure. At the RFC-XX-M experiment where there was a $B=0$ region in the cusp (see Sec.~\ref{sec:RFC}), a hollow density profile in the radial direction was indeed experimentally found~\cite{Baker1984}. As interchange modes are driven by pressure gradients, one must therefore study not only the region near the outer plasma edge but also the inner region of the Novatron. 

There are several theoretical approaches to interchange stability, three of which will be considered here. First, we will use a slightly modified version of one of the most established criteria for anisotropic plasmas introduced by Rosenbluth \& Longmire, relying on particle orbit theory. We modify the criterion slightly to make it applicable also to the inner Novatron plasma region. This is the Modified Rosenbluth \& Longmire (MRL) criterion. Secondly, we will present two new criteria, the Generalized Rosenbluth \& Longmire Interchange (GRLI) criterion, based on their MHD interchange criterion, and the Chew-Goldberger-Low Interchange (CGLI) criterion. These two criteria are both derived within fluid models, but in contrast to the MRL criterion, the GRLI criterion, which based on a collisional model, does not require the adiabatic invariants $\mu =  v_{\perp}^2/2B$ and $J = \oint v_{\parallel}d\ell$ to be conserved. The CGLI criterion, on the other hand, derived using the Chew-Goldberger-Low (CGL)~\cite{CGL} equations, employs separate equations of state for $p_\parallel$ and $p_\perp$, which are equivalent to conservation of the two adiabatic invariants $\mu$ and $J$, in order to be an accurate description of collisionless anisotropic plasmas It should be noted that MHD models become less valid near the $B=0$ regions since the locality assumption will be violated due to the large Larmor radius. Consequently, the results relating to these criteria are less reliable in these regions.

All three criteria require that plasma pressure profiles are specified. In Sec.~\ref{sec:pressure_profiles}, two analytical families of pressure profiles, the Taylor and the Cutler profiles, are presented. The interchange criteria have been evaluated for both these sets of pressure profiles.

In the following sections we will assume low-$\beta$ equilibria for which induced (perturbed) magnetic fields are negligible. Thus, they can be expressed as functions of the vacuum magnetic field. Furthermore, there is no change in magnetic energy if interchange modes develop. 

\subsection{Anisotropic Novatron equilibria} \label{sec:pressure_profiles}

The pressure profiles under consideration are those typically found in minimum-B devices and classic mirrors. They are given explicitly in analytical form. 

To start, we can define the parallel pressure profile to be dependent on the external magnetic field $B$ and flux $\psi$ as follows,
\begin{align}
    p_{\parallel}(\psi, B) & = A(\psi) p_{\parallel}(B).
\end{align}
Here, the pressure profile has been split into two factors: $A(\psi)$, which allows us to limit the pressure to a certain set of flux-lines, and the magnetic field term $p_{\parallel}(B)$. 
Once the parallel pressure $p_{\parallel}$ is defined, the perpendicular pressure $p_{\perp}$ can be obtained from the relation~\cite{Grad1967},
\begin{align}
    p_{\perp} = p_{\parallel} - B\frac{d p_{\parallel}}{d B}.
    \label{eq:GC}
\end{align}
$A(\psi)$ is defined as
\begin{align} 
    A(\psi) = \frac{1}{\psi_n^2} (\psi_+ - \psi) (\psi - \psi_-), \label{eq:FischerKillen} 
\end{align} 
where $\psi_+$ ($\psi_-$) is the highest (lowest) value of $\psi$ for which we model the plasma pressure $p$. Elsewhere $p = 0$. Here, $\psi_n = \max(\psi_+, \psi_-)$. Eq.~(\ref{eq:FischerKillen}) is a normalized version seen in~\cite{Fisher:1971}. We denote this as the "Killeen" $A(\psi)$ form. It allows us to model the plasma pressure as slightly hollow, that is with the pressure decreasing towards the z axis.


\paragraph{Cutler pressure profiles}

\begin{figure}
\includegraphics[width=78mm]{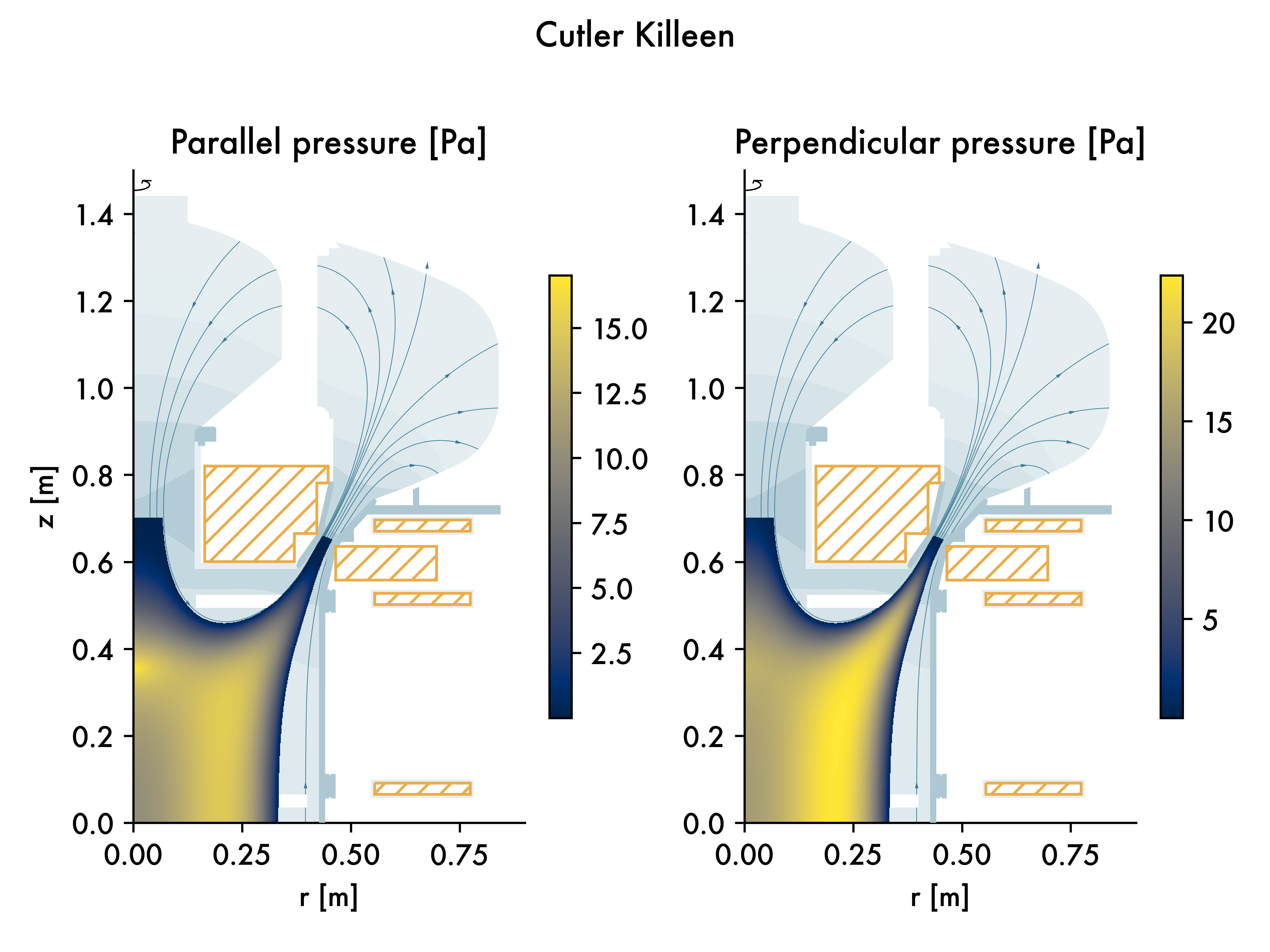}
\caption{Cutler (with the "Killeen" $A(\psi)$ function) parallel (left) and perpendicular (right) pressure distributions for the central cell. $\psi_{+}=3.5 \cdot 10^{-3}$, $\psi_{-}=-9.2\cdot 10^{-4}$, $B_0=0.43$ and $C=65.4$. $C$ is chosen such that the volume integrated $\beta$ value is 1\%.} \label{fig:pressure_profiles_Culter_Killeen}
\end{figure}

Interesting pressure profiles are derived by Cutler~\cite{anderson_calculations_1982}. These are obtained as a fit from a transport code, and may potentially model experimental low pressure profiles more accurately;
\begin{align}
    p_{\parallel}(B) & = C(1 - (B/B_{1}) + (B/B_{1})\text{ln}(B/B_{1})),        \\
    p_{\perp}(B) & = C(1-(B/B_{1})),
    \label{eq:cutler}
\end{align}
if $B \leq B_{1}$, else $p_{\parallel} = p_{\perp} = 0$. Fig.~\ref{fig:pressure_profiles_Culter_Killeen} shows the Cutler pressure profiles for parallel and perpendicular pressures, using the Killeen $A(\psi)$ form. In this figure, the $\psi_{+}$ ($\psi_{-}$) parameter is taken as the $\psi$ value of the outermost (innermost) field line, as defined by the outer (inner) limiter. The $B_{1}$ parameter is set to the magnetic field strength value $B$ at the outermost field line in the mirror throat.

\paragraph{Taylor pressure profiles}

We have also considered another well known set of pressure profiles derived by J.B. Taylor~\cite{Taylor:1963}. For a minimum-B device they are given by
\begin{align}
    p_{\parallel}(B) & = CB(B_{1} - B)^M,        \\
    p_{\perp}(B)     & = CMB^2(B_{1} - B)^{M-1},
    \label{eq:taylor}
\end{align}
if $B \leq B_{1}$, else $p_{\parallel} = p_{\perp} = 0$.

The parallel pressure is here at its minimum at the mirror point, and increases as the magnetic field decreases. This continues to a point of minimum-B. Beyond this minimum-B point the pressure decreases to zero. Lacking any physical explanation for this behavior, we consider this profile unphysical for the Novatron. In addition, our PIC-code simulations (to be published in a parallel paper) strongly support Cutler-like profiles in favor of Taylor-like. Lastly, the stability patterns, which will be presented in Sec.~\ref{sec:num}, are almost identical for Taylor and Cutler profiles. We therefore do not present our results for the Taylor profiles.


\subsection{MHD interchange stability} \label{sec:interchange}

In this section, the classic anisotropic Rosenbluth \& Longmire~\cite{Rosenbluth1957} criterion for interchange stability, based on particle orbit theory, is modified to be applicable to the Novatron's potentially slightly hollow pressure profile. Additionally, two novel fluid interchange stability criteria for anisotropic plasmas are derived, based on anisotropic ideal MHD, and the double-adiabatic, Chew-Goldberger-Low (CGL) models, respectively. 

\subsubsection{Modified Rosenbluth \& Longmire (MRL) criterion}

Rosenbluth \& Longmire~\cite{Rosenbluth1957} derived two interchange stability criteria at low $\beta$; one for isotropic plasmas, based on ideal MHD, and one for anisotropic plasmas, based on particle orbit theory. The latter criterion assumes that the magnetic moment $\mu$ and the longitudinal adiabatic invariant $J$ are both conserved. In Sec.~\ref{sec:J_mu_Novatron} the validity of these assumptions for the Novatron is investigated numerically. The original Rosenbluth \& Longmire criterion reads 
\begin{align}
    \int{\frac{p_{tot}}{RrB^2}}d\ell > 0, \label{eq:RnL}
\end{align}

\noindent for stability, where $p_{tot} = p_\parallel + p_\perp$, $r$ is the radial coordinate and $R$ is the field line curvature radius. In this formulation, $R$ is defined as positive ``if the center of curvature lies outside the plasma". If $R$ is positive everywhere along the field line, the integral will be positive. This definition assumes the pressure gradient always being directed in to the plasma. 

Consider the curvature vector $\kappa$, which is directed towards the center of curvature, irrespective of the plasma's location. Thus, according to Rosenbluth \& Longmire, positive $R$ should be interpreted as the pressure gradient being directed anti-parallel to $\kappa$. Hence, using this definition instead, we can apply the criterion to cases where the direction of the pressure gradient points inwards. We can then apply the Rosenbluth \& Longmire criterion also on the inner side of the plasma in the Novatron. The criterion then becomes:
\begin{align}
\int{-\text{sign(}\vec{\kappa} \cdot \nabla p_{tot}) \frac{p_{tot} |\vec{\kappa}|}{rB^2}} d \ell > 0. \label{eq:RnL_criterion}
\end{align}

We denote this criterion the Modified Rosenbluth \& Longmire (MRL) criterion. Numerical results for Novatron geometry are presented in Sec.~\ref{sec:num}. 

\subsubsection{Generalized Rosenbluth \& Longmire Interchange (GRLI) criterion} \label{sec:GRLI}

We will now derive a novel interchange stability criterion, based on anisotropic ideal MHD. This generalized Rosenbluth \& Longmire criterion does not require the adiabatic moments $\mu$ and $J$ to be conserved, as is the case for the (particle orbit theory-based) MRL criterion. Furthermore, the criterion takes into account pressure variations along the magnetic field lines.

The ideal MHD model is derived in the collision-dominated limit where particle collision times are shorter than other typical time scales. Further, isotropization times for parallel and perpendicular ion pressures are of similar order as the ion-ion collision time $\tau_{ii}$ in the absence of anisotropic driving forces. Consequently, the plasma is considered isotropic in ideal MHD. However, to neglect resistivity, ideal MHD must assume $\tau_{ii}$ to be finite. Thus, anisotropic effects generated on short time scales, for example by strong NBI and RF heating, or by mirror confinement, could then consistently be included in the model without violating other assumptions~\cite{Shi, Brunetti}. 

To find a criterion for interchange stability we will, along the lines of Rosenbluth \& Longmire, investigate the circumstances under which it is energetically beneficial for the plasma to interchange flux tubes. A derivation of the stability criterion now follows; more details are found in the Appendix.

The (adiabatic) ideal MHD energy equation is
\begin{align}
    \frac{d}{dt}\Big(p \rho^{-\gamma}\Big) = 0, \label{eq:MHD}
\end{align}
where $\rho$ is the charge density. After integration we may write, where $'$ denotes a time $t=t'$ and $''$ denotes a later time $t=t''$,
\begin{align}
\frac{p'}{\rho'^{\gamma}} = \frac{p''}{\rho''^{\gamma}}. \label{eq:MHD_time} 
\end{align}

Consider now two infinitesimally adjacent magnetic flux tubes, denoted by subscript 1 and 2, reaching all the way to the expander regions of the device. At time $t=t'$ flux tube 1 is at position A and flux tube 2 is at position B. We assume that at time $t=t''$ the two flux tubes are interchanged, so that flux tube 1 is at position B while flux tube 2 is at position A. Assuming negligible induced fields ($\nabla \times \textbf B = \textbf 0$), the flux in the two flux tubes is the same, that is $\phi_1 = \phi_2$ in order to neglect the potentially stabilizing effect of field line bending, as shown by Rosenbluth \& Longmire~\cite{Rosenbluth1957}. As a consequence, the change in magnetic energy induced by the interchange is zero for both flux tubes. The perturbation in the radial direction however corresponds to a change in the magnetic flux function $\delta \psi$, where $\psi$ is associated with the total flux as a function of radius through the symmetry plane. 

We express local changes in terms of two infinitesimal cylindrical sections of the flux tubes with area $S$, height $d\ell$, volume $v$ and pressure $p$. During the interchange, these infinitesimal flux tubes are assumed to interchange volumes, so that
\begin{align}
v_1'' = v_2', \label{eq:volume_1} \\
v_2''= v_1', \label{eq:volume_2}
\end{align}

\noindent while the change in pressure is governed by Eq. (\ref{eq:MHD}). The pressure may vary along the flux tube. The induced field is taken to be negligible so that 
\begin{align}
B_1'' = B_2', \label{eq:no_line_b_1} \\
B_2'' = B_1'. \label{eq:no_line_b_2}
\end{align}

The internal (or material) plasma energy density is
\begin{align}
    E_p = \frac {3}{2}(n_ek_BT_e + n_ik_BT_i) = 3nk_BT = \frac {3}{2} p,
\end{align}
for equal ion and electron densities and temperatures, with $p = (p_{\parallel} + p_{\perp})/2$. The material energy of an infinitesimal flux tube with height $d\ell$ can thus be written 
\begin{align}
    dE = \frac {3}{2} pv. 
\end{align}
 To arrive at a criterion for interchange stability, we would like to express the total change in energy associated with the interchange of the two entire flux tubes. Since the change in magnetic energy induced by the interchange is zero, this total change in energy is equal to the total change in material energy, $\Delta E$.

 We will begin by expressing the local changes where the change in energy for the two infinitesimal flux tubes $\Delta (dE) = \Delta (dE_1) +\Delta (dE_2)$ is then related to the stability criterion as follows:

 \begin{align}
	\Delta E = \Delta \int dE = \int \Delta (dE) > 0  ,
\end{align}

 and the integral sums up the changes in material energy of the infinitesimal flux cylinders along a flux tube.
 We write, omitting the factor $3/2$,
\begin{align}
    & \Delta( dE_{1}) = p_1''v_1'' - p_1'v_1' = p_1''v_2' - p_1'v_1', \label{eq:deltaE1} \\
    & \Delta (dE_{2}) = p_2''v_2'' - p_2'v_2' = p_2''v_1' - p_2'v_2'. \label{eq:deltaE2}
\end{align}
Let us now rewrite $p_1''$ using Eqs. (\ref{eq:MHD_time}) and (\ref{eq:volume_1}):

\begin{align}
p_1'' = p_1' \Big( \frac{\rho_1''}{\rho_1'} \Big)^{\gamma} = p'_1 \Big( \frac{v_1'}{v_1''} \Big)^{\gamma} = p'_1 \Big( \frac{v_1'}{v_2'} \Big)^{\gamma}, \label{eq:p_1_bis}
\end{align}



and similarly for $p_2''$ obtaining 
\begin{align}
p_2'' = p'_2 \Big( \frac{v_2'}{v_1'} \Big)^{\gamma}. \label{eq:p_2_bis}
\end{align}
Insertion of Eqs. (\ref{eq:p_1_bis}) and (\ref{eq:p_2_bis}) into Eqs. (\ref{eq:deltaE1}) and (\ref{eq:deltaE2}) yields
\begin{align}
& \Delta( dE_{1}) = p'_1 \Big( \frac{v_1'}{v_2'} \Big)^{\gamma} v_2' - p_1'v_1', \label{eq:deltaE1_2} \\
& \Delta( dE_{2}) =  p'_2 \Big( \frac{v_2'}{v_1'} \Big)^{\gamma} v_1' - p_2'v_2'. \label{eq:deltaE2_2}
\end{align}

Having expressed $\Delta( dE_{i})$ in terms of quantities at time $t=t'$ only, we can drop the $'$ superscript. Next, we rewrite the pressure, volume and magnetic field strength for flux tube 2 in terms of those for flux tube 1 (at time $t=t'$) and obtain

\begin{align}
    p_2 = p_1 + \delta p, ~~~ v_2 = v_1 + \delta v. \label{eq:p_v_delta}
\end{align}

Inserting this into Eqs.~(\ref{eq:deltaE1_2}) and~(\ref{eq:deltaE2_2}), keeping terms up to 2nd order and setting $p_1=p$ and $v_1=v$, we can write (see Appendix for details) 
\begin{align}
    \Delta (dE) &= \Delta (dE_1) +\Delta (dE_2) \\
    &= (\gamma-1) \Bigg(\delta p + \gamma p \frac{\delta v}{v}\Bigg)\delta v  \label{eq:deltaEp}
\end{align} 

 As mentioned above, to arrive at a criterion, we need to add the contributions from all infinitesimal flux tubes. We note that the flux $\phi = SB$ is constant along the flux tube so that
\begin{align}
     v = A d\ell = \phi \frac{d\ell}{B}. \label{eq:phi}
\end{align}

We obtain (for details see Appendix): 

\begin{align}
	\int \delta \Bigg(\frac {1}{B}\Bigg) B^{2 \gamma} \delta \Bigg( \frac {p_{\parallel} + p_{\perp}} {2B^{2 \gamma}}\Bigg) d\ell > 0  .
\end{align}

This is the Generalized Rosenbluth \& Longmire Interchange (GRLI) criterion. Here the functions, on which $\delta$ operate, are non-constant along the field lines. For practical application, we may assume the paraxial limit in which $|d/dr| \gg |d/dz|$ and introduce the flux coordinate $\psi$, arriving at the approximate stability condition
\begin{align}
    \int \frac{\partial}{\partial \psi}\Bigg(\frac{1}{B}\Bigg)B^{2\gamma}\frac{\partial}{\partial \psi}\Bigg(\frac{p_{\parallel} + p_{\perp}}{2B^{2\gamma}}\Bigg) d\ell > 0. \label{eq:GRLI_criterion}
 \end{align}

Clearly, a sufficient condition for stability is obtained if the integrand is everywhere positive, that is
\begin{equation}
	\frac {\partial}{\partial \psi} \Bigg(\frac {1} {B}\Bigg) \ \frac {\partial}{\partial \psi} \Bigg(\frac {p_\parallel + p_\perp} {2B^{2 \gamma}}\Bigg) > 0.
\end{equation}

For stability in a magnetic well, with $(\partial/\partial \psi)(1/B)< 0$, $p$ should thus decrease with $\psi$ or increase slower than $B^{2 \gamma}$. The marginal, sufficient condition becomes
\begin{equation}
	\frac {\partial}{\partial \psi} \Bigg(\frac {p_\parallel + p_\perp} {2B^{2 \gamma}}\Bigg) = 0 ,
\end{equation}

\noindent or 
\begin{equation}
	p_\parallel + p_\perp = C  {B^{2 \gamma}},
\end{equation}

\noindent where $C$ is a constant.

One should note that the criterion Eq. (\ref{eq:GRLI_criterion}) allows for a stronger variation of $p$ with respect to $\psi$, and thus a more hollow pressure profile.


\subsubsection{Chew-Goldberger-Low Interchange (CGLI) criterion} \label{sec:CGLI}

The interchange stability criterion (\ref{eq:GRLI_criterion}) is general, and will be evaluated for low-$\beta$ plasmas in the Novatron magnetic field in Sec.~\ref{sec:num}. Since the criterion is based on collisional MHD and fusion plasmas are typically collisionless, we will now, for comparison, derive a corresponding interchange stability criterion based on the collisionless Chew-Goldberger-Low (CGL) fluid model~\cite{CGL, Hunana:2019, Kaur:2019, Le:2016}. Some details of the derivation of this Chew-Goldberger-Low Interchange (CGLI) criterion can be found in the Appendix. 
 
The CGL model assumes vanishing heat flux, and features the following adiabatic, anisotropic equations of state:
\begin{align}
    &\frac{d}{dt}\Bigg(\frac{p_{\parallel}B^2}{\rho^3}\Bigg) = 0, \\
    &\frac{d}{dt}\Bigg(\frac{p_{\perp}}{\rho B}\Bigg) = 0, 
\end{align}
or, using similar terminology as in Sec.~\ref{sec:GRLI}, 
\begin{align}
\frac{p_{\parallel}'B'^2}{\rho'^3}=\frac{p_{\parallel}''B''^2}{\rho''^3}, \label{eq:dad_1} \\
\frac{p_{\perp}'}{\rho' B'}=\frac{p_{\perp}''}{\rho'' B''},  \label{eq:dad_2}
\end{align}
 valid for any infinitesimal plasma fluid element. Consider two adjacent flux tubes, as defined in Sec.~\ref{sec:GRLI}, but where the pressures are now governed by (\ref{eq:dad_1}) and (\ref{eq:dad_2}). Using Eqs. (\ref{eq:dad_1}) and (\ref{eq:dad_2}), valid for both flux tube 1 and 2 separately, we obtain
\begin{align}
    p_1'' &= \frac{1}{2}(p_{\parallel, 1}'' + p_{\perp,1}'') \nonumber \\
    &=\frac{1}{2}\Bigg(\frac{p_{\parallel,1}'B_1'^2v_1'^3}{B_1''^2v_1''^3} + \frac{p_{\perp 1}'v_1'B_1''}{v_1''B_1'}\Bigg), \label{eq:p_bis_1}
\end{align}
and a similar expression for $p_2''$. We now (see Appendix for details) use Eqs.~(\ref{eq:volume_1}) and (\ref{eq:no_line_b_1}) in Eq.~(\ref{eq:p_bis_1}), and similarly for the $p_2''$ expression, and then substitute these into Eqs.~(\ref{eq:deltaE1}) and (\ref{eq:deltaE2}) to obtain
\begin{align}
  \Delta (dE) &= \Delta (dE_1) +\Delta (dE_2) \nonumber \\
    &=\frac{1}{2}\Bigg[\Bigg(\frac{p_{\parallel, 1}'B_1'^2v_1'^3}{B_2'^2v_2'^3} + 
     \frac{p_{\perp, 1}'v_1'B_2'}{v_2'B_1'}\Bigg)v_2' \nonumber \\
     &-(p_{\parallel 1}' + p_{\perp 1}')v_1' \nonumber \\
    &+\Bigg(\frac{p_{\parallel 2}'B_2'^2v_2'^3}{B_1'^2v_1'^3} + 
    \frac{p_{\perp 2}'v_2'B_1'}{v_1'B_2'}\Bigg)v_1' \nonumber \\
    &- (p_{\parallel 2}' + p_{\perp 2}')v_2'\Bigg]. \label{eq:delta_E_prim_only}
\end{align}


Having expressed $\Delta (dE) $ in terms of quantities at time $t=t'$ only, we can drop the $'$ superscript. Next, the pressures, volumes, and magnetic field strength for flux tube 2 are rewritten in terms of those for flux tube 1 (both at time $t=t'$):
\begin{align}
    p_{\parallel 2} = p_{\parallel 1} + \delta p_{\parallel}, \label{eq:p_papa_delta} \\
    p_{\perp 2} = p_{\perp 1} +  \delta p_{\perp}, \label{eq:p_perp_delta} \\
    v_{2} = v_{1} + \delta v, \label{eq:v_delta} \\
    B_{2} = B_{1} + \delta B. \label{eq:B_delta}
\end{align}
Substituting these equations into Eq.~(\ref{eq:delta_E_prim_only}), setting $p=p_1$, and keeping terms up to 2nd order we obtain after some algebra (see Appendix for details):

 \begin{align}
  \Delta E \propto \bigintsss\delta \Bigg(\frac{1}{B}\Bigg) \Bigg[B^4\delta\Bigg(\frac{p_{\parallel}}{B^4}\Bigg)  \nonumber \\
   + \frac{B^3}{2}\delta\Bigg(\frac{p_{\perp}}{B^3}\Bigg)\Bigg]d\ell > 0. \label{eq:CGL_criterion}
\end{align}

 We denote this criterion as the Chew-Goldberger-Low Interchange (CGLI) criterion. In the paraxial limit, the stability condition is approximated by

\begin{align}
    \bigintsss \frac{\partial}{\partial \psi} \Bigg(\frac{1}{B}\Bigg) \Bigg[B^4\frac{\partial}{\partial \psi}\Bigg(\frac{p_{\parallel}}{B^4}\Bigg) + \nonumber \\
   \frac{B^3}{2}\frac{\partial}{\partial \psi}\Bigg(\frac{p_{\perp}}{B^3}\Bigg)\Bigg]d\ell > 0. \label{eq:CGL_criterion}
\end{align}

\subsubsection{Validity of the adiabatic invariance assumption} \label{sec:J_mu_Novatron}

 the MRL and the CGLI interchange stability criteria rely on conservation of the adiabatic invariants $\mu$ and $J$. Because of the presence of weak field regions in the Novatron vacuum field, we have evaluated the variation of $\mu$ and $J$ in the plasma domain. For comparison, we have also evaluated the variation of these invariants for a classic mirror. The results are given in Figs.~\ref{fig:invariance_m67_001} and  \ref{fig:invariance_m67_002}. 

\begin{figure}
\includegraphics[width=78mm]{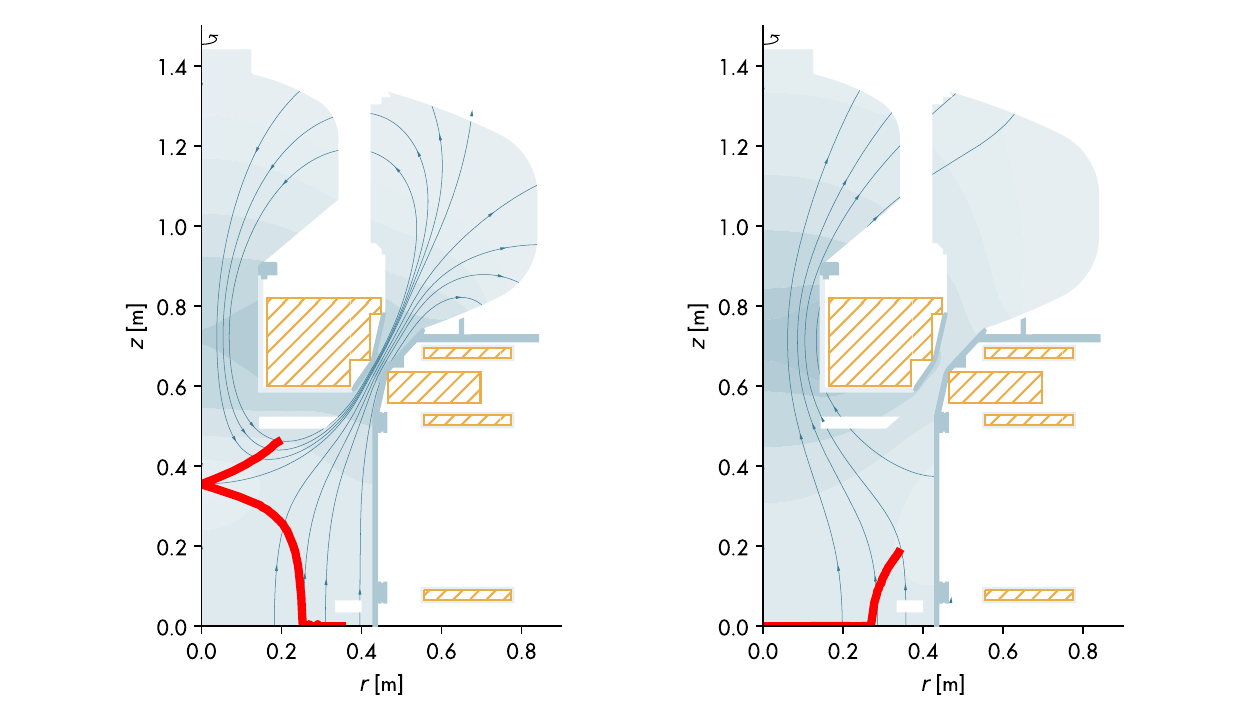}
\caption{The line of minimum $B$ per field line for the Novatron vacuum field (left) and classic mirror vacuum field (right).}\label{fig:minb}
\end{figure}

\begin{figure}
\includegraphics[width=78mm]{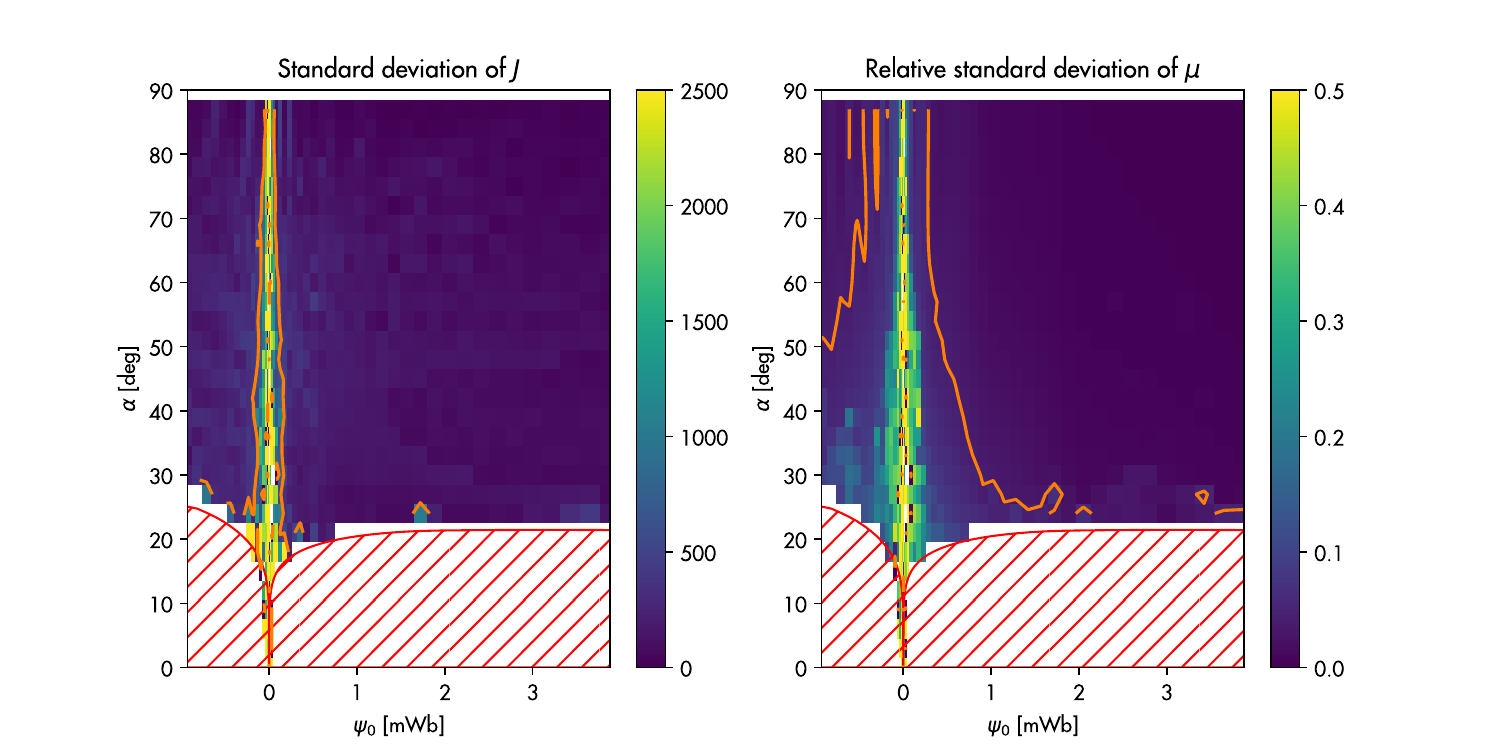}
\caption{The standard deviation of $J$ (left) and the relative standard deviation of $\mu$ (right) for particles in a Novatron vacuum field. The hatched red area at low values of $\alpha$ represents the loss cone.}\label{fig:invariance_m67_001}
\end{figure}

\begin{figure}
\includegraphics[width=78mm]{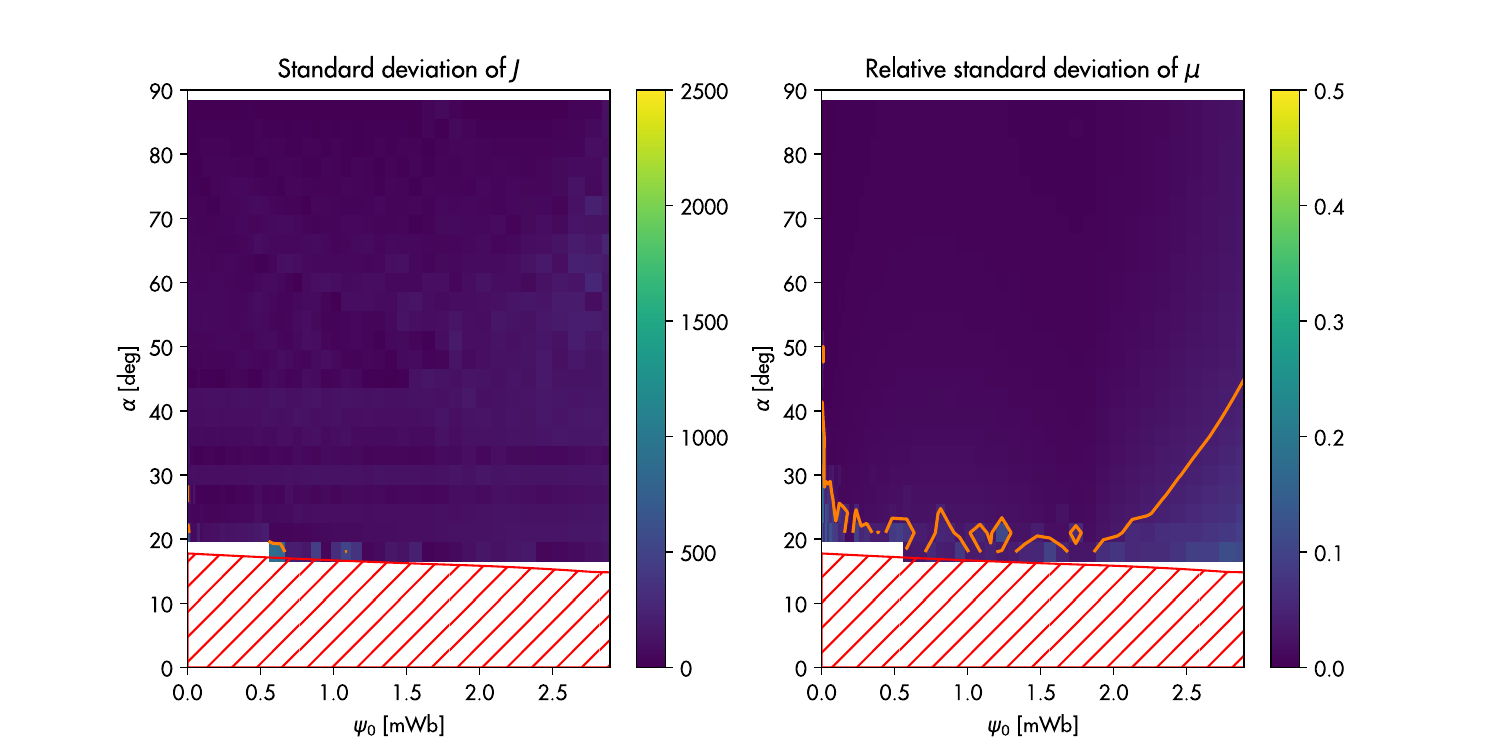}
\caption{As Fig.~\ref{fig:invariance_m67_001} but for a classic mirror vacuum field.}\label{fig:invariance_m67_002}
\end{figure}

For each vacuum field, we tracked 10 eV hydrogen ions under the Lorentz force for 1 s. We chose 50 different starting points from a line of minimum $\textbf B$ per field line, with higher resolution near $\phi=0$, see Fig.~\ref{fig:minb}. For each starting point, pitch angles, $\alpha$, were selected as 30 equidistant points on the interval $ [0,\pi /2]$. The values of the two adiabatic invariants were recorded each time the particles passed the minimum $\textbf B$ line in the direction parallel to the $\textbf B$ field (once per bounce), and the mean and standard deviations were calculated. For the magnetic moment $\mu$, the standard deviation relative to the mean is presented, whereas for the action integral $J$ (which has a mean close to zero) the standard deviation is shown.

To aid the comparison between the two vacuum field geometries, we show an isoline at a single level: $\sigma_\mu / \overline{\mu} = 0.025$ for the magnetic moment and $\sigma_J = 500$ for the action integral. These limit values could have been chosen differently without altering the qualitative conclusions.

If we consider the areas in the $\psi-\alpha$ space below these levels as indicating a region where the invariance of $\mu$ and $J$ is respected, we see that all field lines in the Novatron field outside $\psi > 1 \mathrm{\,mWb}$ conserve the magnetic moment. On the other hand there is a narrow region for field lines with $\psi$ close to zero, where there is no pitch angle for which the invariants are conserved. Fortunately, for both kinds of pressure profiles of Sec.~\ref{sec:pressure_profiles}, the majority of the plasma is located at $\psi$ values close to the maximum pressure, at around $r = 0.2 \mathrm{\, m}$ (see Fig. \ref{fig:pressure_profiles_Culter_Killeen}), corresponding to $\psi = 1.22 \mathrm{\, mWb}$, where both adiabatic invariants are conserved. 

In the classic mirror vacuum field, $J$ is conserved outside the loss cone. In the proximity of the loss cone boundary however, some particles escape due to pitch angle variation. Similarly, $\mu$ is conserved for $\psi < 1.8 \mathrm{\, mWb}$, except near the loss cone. For field lines with $\psi > 1.8 \mathrm{\, mWb} $, $\mu$ is not conserved for low pitch angles, corresponding to regions where the minimum $\textbf B$ line m the symmetry plane ($r > 0.28 \mathrm{\, m}$), invalidating the paraxial approximation.


\subsubsection{Numerical results for Novatron interchange stability} \label{sec:num}

\begin{figure}
\includegraphics[width=78mm]{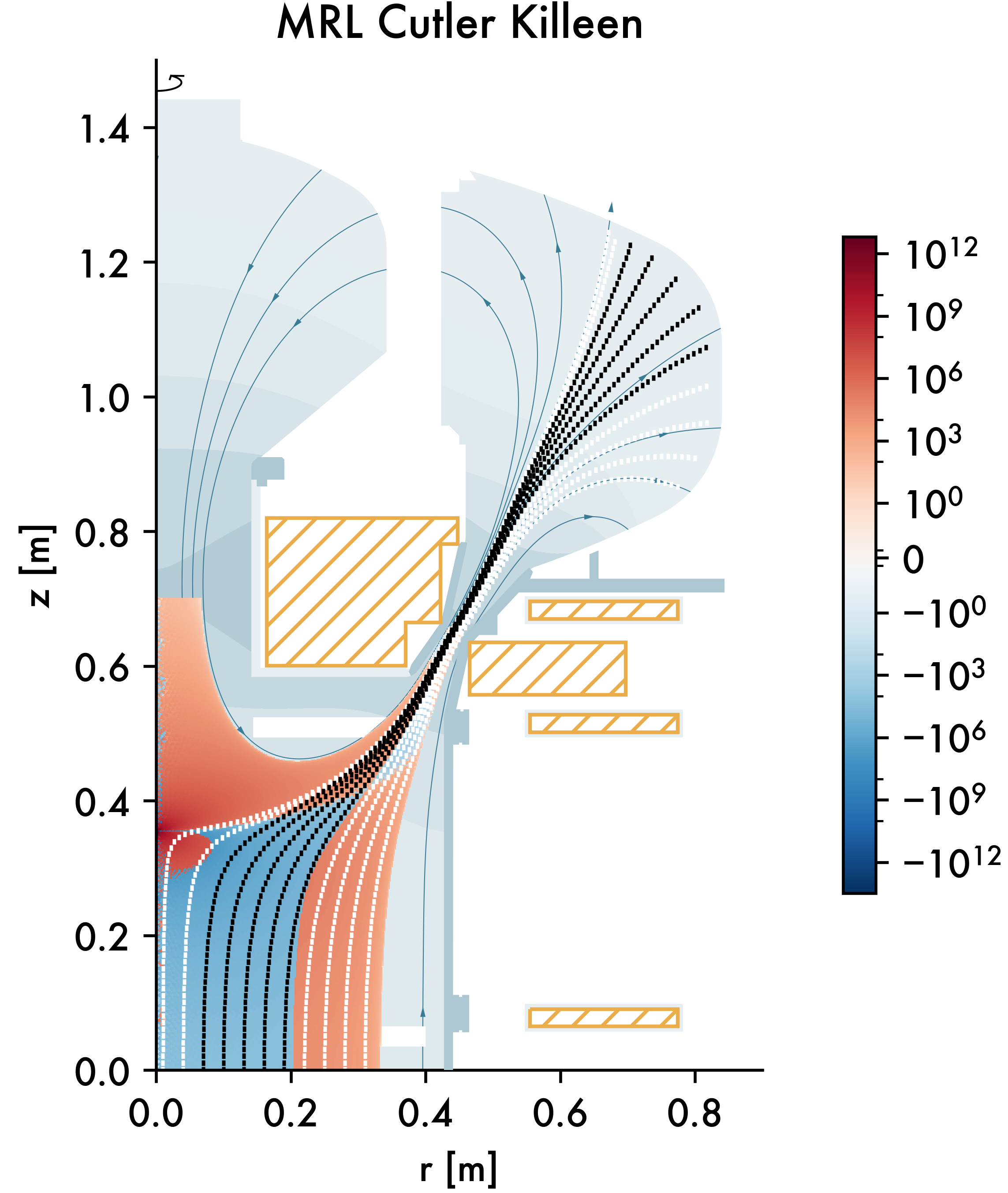}
\caption{Integrand for the MRL criterion assuming a Cutler pressure profile (with the "Killeen" {$A(\psi)$} form) superimposed on the Novatron chamber. The integrand, with values according to the symlog color bar, is evaluated only where the pressure $>0$. The white (black) dotted lines represent flux tubes, or field lines, which are stable (unstable) against interchange modes according to the MRL criterion integral.}\label{fig:interchange_criterion_MRL_Cutler_Killeen}
\end{figure}

\begin{figure}
\includegraphics[width=78mm]{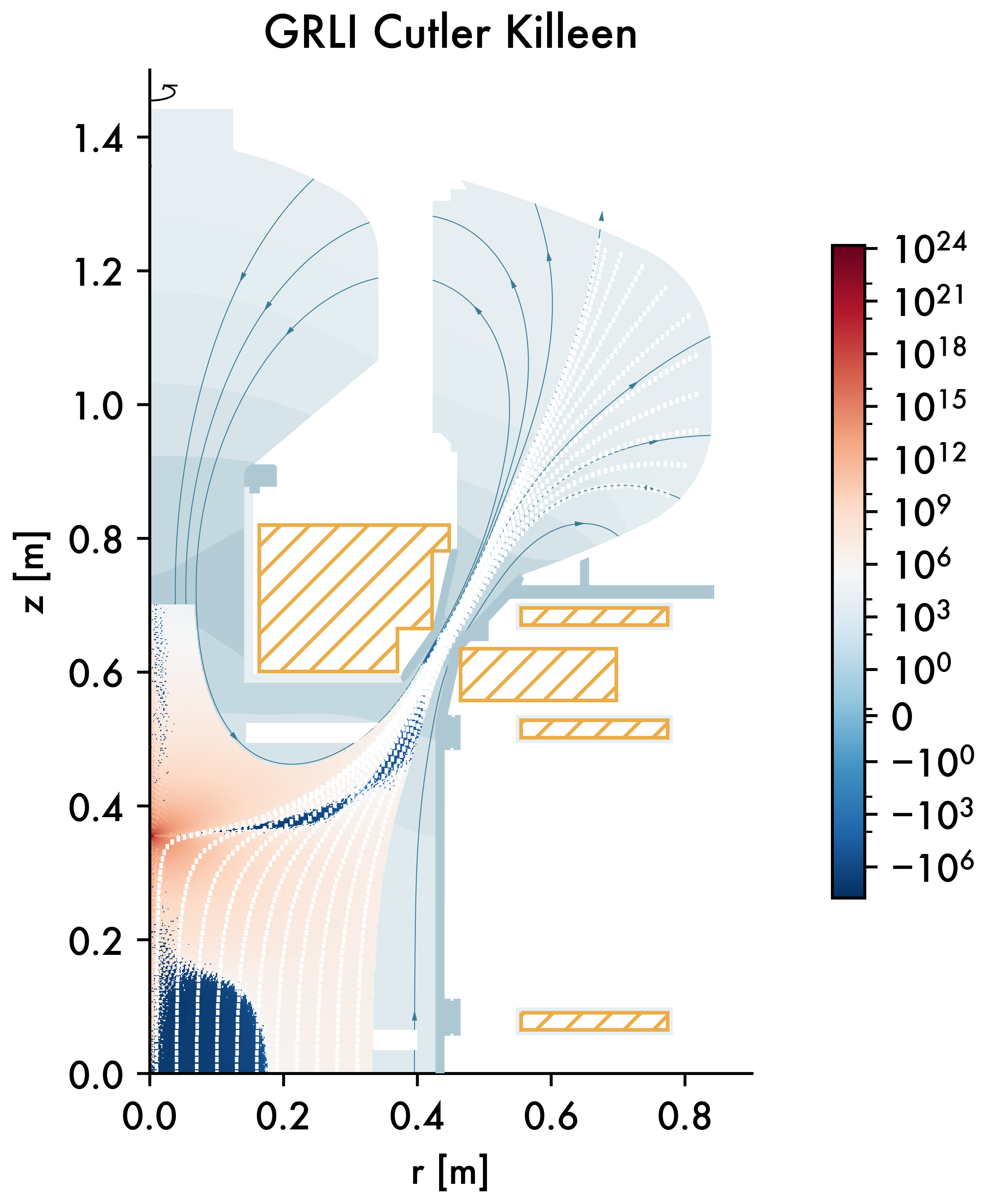}
\caption{Same as in Fig.~\ref{fig:interchange_criterion_MRL_Cutler_Killeen} but for the GRLI criterion .}\label{fig:interchange_criterion_GRLI_Cutler_Killeen}
\end{figure}

\begin{figure}
\includegraphics[width=78mm]{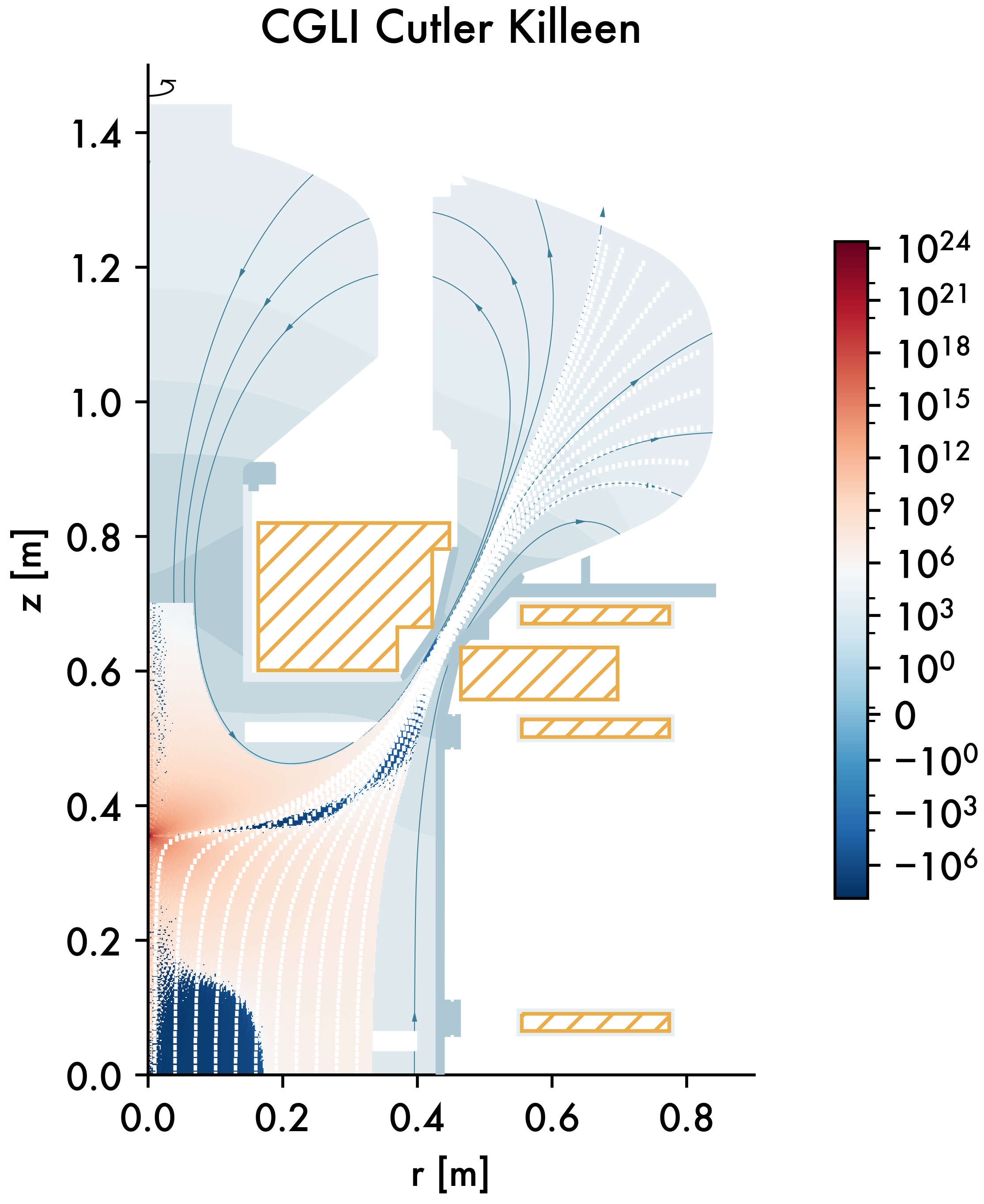}
\caption{Same as in Fig.~\ref{fig:interchange_criterion_MRL_Cutler_Killeen} but for theGLI criterion.}\label{fig:interchange_criterion_GRLI_and_CGL_Cutler_Killeen} \label{fig:interchange_criterion_CGLI_Cutler_Killeen}
\end{figure}

The MRL, GRLI and CGLI stability criteria have been evaluated in a Novatron magnetic field geometry for the analytic anisotropic Cutler pressure profiles of Sec.~\ref{sec:pressure_profiles} for $\gamma=5/3$, as seen in Figs.~\ref{fig:interchange_criterion_MRL_Cutler_Killeen} - \ref{fig:interchange_criterion_CGLI_Cutler_Killeen}. The integrals are evaluated for a set of field lines and superimposed on the local values of the integrands of the criteria. The integrands are shown in order to expose which regions contribute positively and which contribute negatively to the integrals. The field lines shown are only in the inner plasma domain since, as seen in the figures, the integrand is positive in the whole outer plasma domain for all three criteria. The expanders were excluded but are assumed to contribute positively to the integral (and thus strengthen stability) in all three criteria since the curvature would here be favorable throughout, assuming pressure profiles approximately proportional to the magnetic field strength. The vacuum field was exported (on a mesh) from Altair Flux 2D simulation software and interpolated using Lagrange polynomials.

The MRL integrand is negative where the pressure gradient is parallel to the curvature vector $\kappa$. Assuming a slightly annular plasma (as is the case of the Cutler equilibrium profile with a Killeen $A(\psi)$ form), there is therefore an inner plasma region where the flux tubes, or field lines, are unstable, see Fig.~\ref{fig:interchange_criterion_MRL_Cutler_Killeen}. For the GRLI and CGLI criteria, there is a substantially smaller region close to the origin, where the integrand runs negative. In contrast to the MRL integrand, however, this negative region is outweighted by the region of positive integrand at higher $z$ values, resulting in positive integrals for all the field lines and thus a completely stable plasma. For all criteria, there is a region close to the mirror throat where the curvature becomes unfavorable, which renders all integrands negative.

It is interesting to note that the integrands of the GRLI and CGLI criteria are close to identical, and that the resulting degree of stability is qualitatively the same. Since the two plasma models, on which the criteria are based, are collisional and collisionless, respectively, this tends to indicate that the MHD approximation works well in this context. Mathematically the resemblance becomes clear by inspecting the exponents of $B$ in both expressions. For the GRLI integrand the exponent is $2\gamma = 10/3$ which lies just between the two exponents (3 and 4) in the CGLI criterion.


\section{Discussion}\label{sec:discussion}

We have, in this article, presented the Novatron concept and its characteristic vacuum magnetic field geometry. The focus has been on demonstrating the influence of the favorable curvature on plasma interchange stability, using multiple magnetohydrodynamic models. The Novatron stability analysis is carried out for low $\beta$. In a forthcoming publication, we will report results from hybrid particle-in-cell (PIC) WarpX simulations of interchange stability of equilibria also at high $\beta$. 

Studies of axial confinement in the Novatron geometry are carried out in parallel, to be published in a separate article. Not only the effect of the large Novatron mirror ratio, but also effects of added tandem cells, employing both electrostatic confinement of ions and ponderomotive confinement of electrons, are investigated. Since the Novatron geometry is fully axially symmetric, neoclassical (radial) losses associated with drift motion are expected to be relatively low. Nevertheless, radial transport of particles and energy in terms of diffusion, microinstabilites and turbulence constitute a next step for Novatron confinement analysis.



\section{Conclusion}\label{sec:conclusion}

Mirror machine concepts hold potential for compact fusion power plants, capable of confining plasmas at high temperatures and high $\beta$ values. Previous mirror machines however, have encountered challenges due to asymmetry resulting from measures taken to eliminate interchange instability, leading to neoclassical losses, and from kinetic instabilities such as DCLC modes. 

The Novatron concept, presented in this paper, is designed to address these issues. t magnetic topology enables interchange stable plasmas, as demonstrated by applying three anisotropic interchange stability criteria using a Cutler pressure profile. Furthermore, the axisymmetric Novatron geometry will eliminate or significantly reduce neoclassical losses. Lastly, DCLC modes are avoided due to the inherent large ratio between plasma and typical Larmor radii.

A first Novatron prototype, N1, is currently under completion at the Alfvén Laboratory, KTH, Sweden. The primary focus of N1 is to experimentally verify interchange and DCLC stability as demonstrated in the present study.

\backmatter

\section*{Statements and declarations}

\subsection*{Conflict of interest}
Novatron Fusion Group is a limited liability commercial company. All authors are fully or partially employed by Novatron Fusion Group and a subset owns shares or options in the company. J.J. is the founder of the Novatron Fusion Group. The authors declare that they have no conflict of interest.


\subsection*{Acknowledgments}

Jan Jäderberg invented the Novatron concept. Katarina Bendtz, Jan Jäderberg, Jan Scheffel and Kristoffer Lindvall drafted the manuscript and performed the literature search for the review sections. Per Niva, Robin Dahlbäck, Rickard Holmberg, Kristoffer Lindvall, Katarina Bendtz and Jan Scheffel contributed to the development of the concept. Jan Scheffel derived the GRLI criterion and Kristoffer Lindvall derived the CGLI criterion. Kristoffer Lindvall and Rickard Holmberg found the anisotropic pressure profiles and applied them to the Novatron configuration. Katarina Bendtz contributed to the formulation of the derivation of the GRLI and the CGLI criteria. Rickard Holmberg made Figs. 1--4. Rickard Holmberg and Johan Lundberg made Figs. 8 --10. Katarina Bendtz made Figs. 5 -- 6. Katarina Bendtz and Kristoffer Lindvall made Figs. 7 and 11 -- 13. All authors critically revised the content of the manuscript.  

The authors wish to acknowledge valuable conversations with Professor Kenneth Fowler, and for bringing the author's attention to relevant literature and experiments. We would further like to express our greatest gratitude towards Dr.\ Arthur Molvik for important input in general and for detailed information about the 2X series experiments, to Professor Shoichi Okamura for reviewing the RFC experiment section and to Professor Mizuki Sakamoto for providing important edits on the text about the GAMMA 10 experiment. We are deeply grateful to Lars Jäderberg for his invaluable discussions, insights, and simulations. The authors would also like to thank Katherine Dunne, Gustaf Mårtensson and Benjamin Verbeek for revising the manuscript. Finally, the would like to acknowledge the work of the whole Novatron team, all contributing to the realization of the Novatron concept.

\bibliography{sn-bibliography}

\onecolumn 
\section{Appendix}

In this Appendix, details of the derivation of the GRLI and CGLI interchange stability conditions are provided. 

\subsection{GRLI}

Inserting Eqs. (\ref{eq:p_v_delta}) into Eqs. (\ref{eq:deltaE1_2}) and (\ref{eq:deltaE2_2}) yields
\begin{align}
\Delta (dE_1) + \Delta (dE_2) &= p_1\Bigg(\frac{v_1}{v_1 + \delta v}\Bigg)^{\gamma} (v_1 + \delta v) \nonumber \\ 
&+ (p_1 + \delta p)\Bigg(\frac{v_1 + \delta v}{v_1}\Bigg)^{\gamma}v_1 \nonumber \\
&-p_1v_1 - (p_1 + \delta p)(v_1 + \delta v). \label{eq:triE}
\end{align}

\begin{figure}[h]
\centering
\includegraphics[width=78mm]{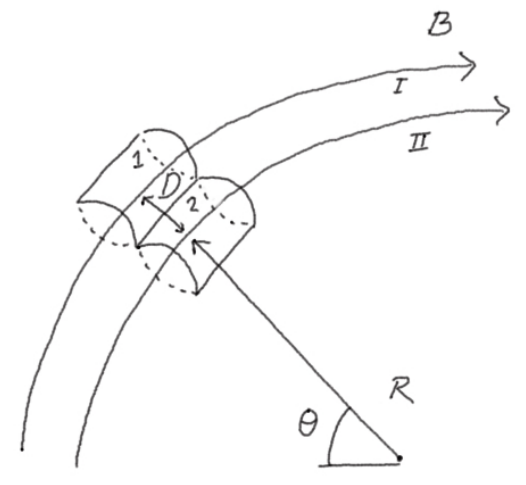}
\caption{Definition of the curvature radius $R$ and distance between tube centers $D$, see the text, drawn on top of two infinitesimal flux tubes 1 and 2. $B$ denotes the magnetic field.}\label{fig:R_D}
\end{figure}

We now set $p_1=p$,  $v_1=v$, and use the approximation

\begin{align}
(1 + x)^{\alpha} \approx 1 + \alpha x + \frac{\alpha (1 - \alpha)}{2}x^2 \text{ for small } x, \label{eq:Taylor_exp}
\end{align}

keeping terms up to 2nd order. We can thus write
\begin{align}
  \Delta (dE) & = \Delta (dE_1) + \Delta (dE_2) 
    = \int(\gamma-1)\delta v\Bigg(\delta p + \gamma p \frac{\delta v}{v}\Bigg) \label{eq:deltaEp}
\end{align} 
Next this expression needs to be simplified. We introduce the variables $D$, the distance between the centers of the infinitesimal tubes, or between the field lines on which the tubes are centered, and $R$, the curvature radius, see Fig. \ref{fig:R_D}. The direction of $R$ is taken to be the same for the two adjacent flux tubes. It follows Eq. (\ref{eq:phi}) that 

\begin{align}
    \delta v = \phi \delta\Bigg(\frac{d\ell}{B}\Bigg) = \phi\Bigg(\frac{\delta(d\ell)}{B} + d\ell\delta\Big(\frac{1}{B}\Big)\Bigg), \label{eq:delta_v}
\end{align}
\begin{align}
    \delta (d\ell) = (d\ell)_2 - (d\ell)_1 = 2\pi d\theta R - 2\pi d\theta(R + D) \\ 
    = -2\pi d\theta D = \Bigg\{d\theta=\frac{(d\ell)_1}{2\pi R}\Bigg\}  
    =-\frac{(d\ell)_1}{R}D. 
\end{align}

Setting $(d\ell)_1 = d\ell$ and using an identity originally derived in Rosenbluth 1957 \cite{Rosenbluth1957}), which holds for $\nabla \times \textbf{B} = \textbf 0$,

\begin{align}
\frac{D}{R}=\frac{\delta B}{B},
\end{align}

gives 

\begin{align}
 \delta (d\ell) = -\frac{\delta B}{B}(d\ell).
\end{align}

Further, using Eq. (\ref{eq:phi}) and Eq. (\ref{eq:delta_v}),
\begin{align}
    \delta v = \phi\Bigg(-\frac{\delta B}{B^2}d\ell - d\ell \frac{\delta B}{B^2}\Bigg)  = -2\phi\frac{\delta B}{B^2}d\ell 
    = -2\phi \delta \Big( \frac{1}{B} \Big) d\ell, \label{eq:delV_delB_id}
\end{align}
\begin{align}
   \frac{\delta v}{v} = -\frac{2\phi\delta B d\ell}{B^2}\frac{B}{\phi d\ell} = -\frac{2\delta B}{B}.
\end{align}

With these expressions Eq. (\ref{eq:deltaEp}) can be simplified. The second factor in Eq. (\ref{eq:deltaEp}) can be written as:

\begin{align}
\delta p + \gamma p \frac{\delta v}{v} &= \delta p - 2p \gamma \frac{\delta B}{B} \\ \nonumber
&= B^{2\gamma} \Big( \frac{\delta p}{B^{2\gamma}} - p\frac{1}{B^{2\gamma + 1}} 2\gamma \delta B \Big) \nonumber \\
&= B^{2\gamma} \delta \Big( \frac{p}{B^{2\gamma}} \Big). 
\end{align}

Now, we can write

\begin{align}
   \Delta (dE) = \phi (\gamma -1) \frac{\partial}{\partial \psi}\Bigg(\frac{1}{B}\Bigg)B^{2\gamma}\frac{\partial}{\partial \psi}\Bigg(\frac{p}{B^{2\gamma}}\Bigg) d\ell
\end{align}

In order to arrive at a criterion, that is an expression for the total change in material energy associated with the flux tube interchange, $\Delta E$, we need to sum over all the infinitesimal flux cylinders along the tubes. 

\begin{align}
	\Delta E = \Delta \int dE = \int \Delta (dE) = \int (\Delta (dE_1) +\Delta (dE_2))> 0  ,
\end{align}

The criterion can thus, omitting the constants $\phi$ and $(\gamma -1)$, be written as

\begin{align}
	\int \delta (\frac {1}{B}) B^{2 \gamma} \delta ( \frac {p_{\parallel} + p_{\perp}} {2B^{2 \gamma}}) d\ell > 0.
\end{align}

\subsection{CGLI}

\subsubsection{From the CGL equations to expressions for $p_1''$ and $p_2''$}

Using Eqs. (\ref{eq:dad_1}) and (\ref{eq:dad_2}) and solving for $p_1''$ and $p_2''$ yields:
\begin{align}
    p_1'' &= \frac{1}{2}(p_{\parallel, 1}'' + p_{\perp,1}'') \nonumber \\
    &=\frac{1}{2}\Bigg(\frac{p_{\parallel,1}'B_1'^2v_1'^3}{B_1''^2v_1''^3} + \frac{p_{\perp 1}'v_1'B_1''}{v_1''B_1'}\Bigg), \label{eqn:p_bis_1}
\end{align}
and 
\begin{align}
     p_2'' &= \frac{1}{2}(p_{\parallel, 2}'' + p_{\perp,2}'') \nonumber \\
    &=\frac{1}{2}\Bigg(\frac{p_{\parallel,2}'B_2'^2v_2'^3}{B_2''^2v_2''^3} + \frac{p_{\perp,2}'v_2'B_2''}{v_2''B_2'}\Bigg). \label{eqn:p_bis_2}
\end{align}
We now use use Eqs. (\ref{eq:volume_1}), (\ref{eq:volume_2}), (\ref{eq:no_line_b_1}) and (\ref{eq:no_line_b_2}) in Eqs. (\ref{eqn:p_bis_1}) and (\ref{eqn:p_bis_2}) to obtain
\begin{align}
    p_1'' &= \frac{1}{2}(p_{\parallel, 1}'' + p_{\perp,1}'') \nonumber \\
    &=\frac{1}{2}\Bigg(\frac{p_{\parallel,1}'B_1'^2v_1'^3}{B_2'^2v_2'^3} + \frac{p_{\perp 1}'v_1'B_2'}{v_2'B_1'}\Bigg), \label{eqn:p_bis_1_simp}
\end{align}

\begin{align}
     p_2'' &= \frac{1}{2}(p_{\parallel, 2}'' + p_{\perp,2}''),¨ \nonumber \\
    &=\frac{1}{2}\Bigg(\frac{p_{\parallel,2}'B_2'^2v_2'^3}{B_1'^2v_1'^3} + \frac{p_{\perp,2}'v_2'B_1'}{v_1'B_2'}\Bigg). \label{eqn:p_bis_2_simp}
\end{align}

\subsubsection{Rewriting $\Delta (dE)$ by expressing differences in the quantities between the flux tubes as perturbations}

Substituting Eqs. (\ref{eq:p_papa_delta}) - (\ref{eq:B_delta}) into Eq. (\ref{eq:delta_E_prim_only}) yields
\begin{align}
    \Delta (dE) &= \frac{1}{2}\Bigg[\frac{p_{\parallel 1}B_1^2v_1^3}{(B_1 + \delta B)^2(v_1 + \delta v)^2} + \frac{p_{\perp 1}v_1(B_1 + \delta B)}{B_1} \nonumber \\
    &- (p_{\parallel 1} + p_{\perp 1})v_1 +\frac{(p_{\parallel 1} + \delta p_{\parallel})(B_1 + \delta B)^2(v_1 + \delta v)^3}{B_1^2v_1^2} \nonumber \\
     &+ \frac{(p_{\perp 1} + \delta p_{\perp})(v_1 + \delta v)B_1}{(B_1 + \delta B)} - (p_{\parallel 1} + \delta p_{\parallel})(v_1 + \delta v) - (p_{\perp 1} + \delta p_{\perp})(v_1 + \delta v)\Bigg].
\end{align}

We set $p_{\parallel 1}=p_{\parallel}$,  $p_{\perp 1}=p_{\perp}
 $, $v_{ 1}=v$, $B_1 = B$, and drop the factor $1/2$. Gathering all $p_{\parallel}$ and $\delta p_{\parallel}$ terms, we define:

\begin{align}
 \Delta (dE_{\parallel}) &= \frac{p_{\parallel}B^2v^3}{(B + \delta B)^2(v + \delta v)^2} - p_{\parallel}v \nonumber + \frac{(p_{\parallel} + \delta p_{\parallel})(B + \delta B )^2 (v + \delta v)^3}{B^2 v^2} +
 (p_{\parallel 1} + \delta p_{\parallel})(v + \delta v).  
 \end{align}
 Next, using Eq. (\ref{eq:Taylor_exp}), neglecting terms of order 3 and higher, 
 we can write

\begin{align}
\Delta (dE_{\parallel}) &= 2\phi d\ell \delta B \frac{1}{B^2} \Bigg( - B^4 \delta \Big(\frac{p_{\parallel}}{B^4}\Big) \Bigg) \nonumber \\ 
&= - 2\phi d\ell \delta B B^2 \Bigg(\delta\Big(\frac{p_{\parallel}}{B^4}\Big) \Bigg) \nonumber \\
&= \Bigg\lbrace\delta \Big(\frac{1}{B} \Big) = -\frac{1}{B^2}\delta B \Bigg\rbrace  \nonumber \\ 
&= 2\phi d\ell B^4 \delta \Big(\frac{1}{B} \Big) \delta \Big(\frac{p_{\parallel}}{B^4}\Big). 
\end{align}


Gathering all the $p_{\perp}$ and $\delta p_{\perp}$ terms, using a similar procedure as for the $p_{\parallel}$ terms we eventually get
\begin{align}
\Delta (dE_{\perp}) &= \phi d \ell \delta \Big(\frac{1}{B}\Big) B^3 \delta \Big(\frac{p_{\perp}}{B^3}\Big).
\end{align}

Thus, our new interchange stability criterion, derived from the CGL double adiabatic equations, becomes
\begin{align}
  \Delta E \propto \bigintsss\delta \Bigg(\frac{1}{B}\Bigg) \Bigg[B^4\delta\Bigg(\frac{p_{\parallel}}{B^4}\Bigg) + 
   \frac{B^3}{2}\delta\Bigg(\frac{p_{\perp}}{B^3}\Bigg)\Bigg]d\ell > 0. \label{eq:CGL_criterion}
\end{align}
where we have dropped the  $2 \phi$ factor. 




\end{document}